\documentclass[11pt]{article}

\usepackage{citesort}
\usepackage{a4wide}
\usepackage{amsmath}
\usepackage{amsthm}
\usepackage{amsfonts,amssymb}
\usepackage{graphicx}

\makeatletter
\@addtoreset{equation}{section}

\makeatother

\def\e{\mathrm{e}}
\def\i{\sqrt{\!-\!1}}
\def\Define{:=}
\def\definE{=:}
\def\nn{\nonumber}
\renewcommand{\leq}{\leqslant}
\renewcommand{\geq}{\geqslant}
\newcommand{\Res}{\mathop{\rm Res}}
\newcommand{\ch}{\mathop{\rm ch}\nolimits}
\newcommand{\vecvar}[1]{\mbox{\boldmath $#1$}}
\newtheorem{thm}{Theorem}[section]
\newtheorem{prop}[thm]{Proposition}
\newtheorem{lm}[thm]{Lemma}
\newtheorem{cor}[thm]{Corollary}

\title
{\Large\textbf{
Completeness of Bethe ansatz for 1D Hubbard model \\[2mm]
with AB-flux through combinatorial formulas \\[2mm]
and exact enumeration of eigenstates
}}

\author{
Akinori Nishino\footnote{E-mail address: %
nishino@gokutan.c.u-tokyo.ac.jp}
and
Tetsuo Deguchi$^{1}$\footnote{E-mail address: %
deguchi@phys.ocha.ac.jp}\\
\,\\
Institute of Physics, University of Tokyo, \\[2mm]
3--8--1 Komaba, Meguro-Ku, Tokyo 153--8902, Japan \\
\,\\
$^{1}$Department of Physics, Ochanomizu University,\\[2mm]
2--1--1 Ohtsuka, Bunkyo-ku, Tokyo, 112--8610, Japan
}

\date{}

\begin{document}

\setlength{\baselineskip}{18pt}

\maketitle
%%%%%%%%%%%%%%%%%%%%%%%%%%%%%%%%%%%%%%%%%%%%%%%%%%%%%%%%%%%%%%%%%%%%%%
\begin{abstract}
\setlength{\baselineskip}{15pt}
For the one-dimensional Hubbard model
with Aharonov-Bohm-type magnetic flux,
we study the relation between its symmetry and
the number of Bethe states.
First we show the existence of solutions for Lieb-Wu 
equations with an arbitrary number of up-spins and one 
down-spin, and exactly count the number of the Bethe states.
The results are consistent with Takahashi's string hypothesis
if the system has the $so(4)$ symmetry.
With the Aharonov-Bohm-type magnetic flux, however, 
the number of Bethe states increases 
and the standard string hypothesis does not hold. 
In fact, the $so(4)$ symmetry reduces to 
the direct sum of charge-$u(1)$ and spin-$sl(2)$ symmetry 
through the change of AB-flux strength.
Next, extending Kirillov's 
approach~\cite{Kirillov_85JSM,Kirillov_87JSM},
we derive two combinatorial formulas from 
the relation among the characters
of $so(4)$- or $\big(u(1)\oplus sl(2)\big)$-modules.
One formula reproduces Essler-Korepin-Schoutens'
combinatorial formula for counting the number of Bethe states
in the $so(4)$-case.
{}From the exact analysis of the Lieb-Wu equations,
we find that another formula corresponds to the spin-$sl(2)$ case.
\end{abstract}

%%%%%%%%%%%%%%%%%%%%%%%%%%%%%%%%%%%%%%%%%%%%%%%%%%%%%%%%%%%%%%%%%%%%%%
\section{Introduction}

Low-dimensional physics has attracted a great interest of 
theoretical and experimental physicists for almost 
a half century~\cite{Korepin-Essler,Mattis,Montorsi}.
Among them, the Bethe ansatz method, which was originally 
developed as a non-perturbative method for diagonalizing 
one-dimensional spin-$\frac{1}{2}$ isotropic Heisenberg 
spin chain~\cite{Bethe_31ZP},
opened a new realm of mathematical physics.
Roughly speaking, Bethe's work consists of three parts:
i) Bethe states are introduced and Bethe equations, 
which are sufficient conditions for 
the Bethe states being eigenstates, are derived;
ii) the existence of solutions for the Bethe equations
are discussed in simple cases, and the general forms of 
solutions are conjectured, which are called string hypothesis; 
iii) under the string hypothesis, the formula for counting 
the number of the Bethe states is provided, which leads to the 
combinatorial completeness of Bethe ansatz.
At present the Bethe ansatz method is applied to various kinds 
of one-dimensional spin chains and strongly correlated 
electron systems~\cite{Mattis,Korepin-Essler,Takahashi_book}.

In Bethe's work, the combinatorial formula for counting 
the number of Bethe states possesses a wealth of mathematical 
implications.
In the case of the isotropic Heisenberg spin chain,
the Bethe states constructed by finite-valued solutions of Bethe 
equations do not produce all the eigenstates.
In fact the system has $sl(2)$ symmetry and the Bethe states 
are $sl(2)$-highest~\cite{Faddeev-Takhtadzhyan_84JSM}.
The eigenstates other than highest weight vectors, 
i.e., $sl(2)$-descendant states, are constructed by applying 
the lowering operator to the Bethe states.
Mathematically, the number of Bethe states is
interpreted as the multiplicity of irreducible components in 
the tensor products of two-dimensional highest
weight $sl(2)$-modules.
Bethe's formula is also extended to the generalized Heisenberg 
spin chains with higher spins or $sl(n)$ symmetry, 
for which the powerful tools such as 
$Q$-systems are introduced~\cite{Kirillov_85JSM,Kirillov_87JSM}.

The application of Bethe ansatz method to the one-dimensional 
Hubbard model was given by Lieb and Wu~\cite{Lieb-Wu_68PRL}.
The Bethe equations for the Hubbard model are often
called Lieb-Wu equations.
Takahashi's string hypothesis asserts that, 
in the thermodynamic limit, the solutions of Lieb-Wu 
equations are approximated by string solutions, 
and the number of Bethe states is estimated under
the hypothesis~\cite{Takahashi_72PTP,Takahashi_74PTP}.
The Hubbard model with even sites has $so(4)$ 
symmetry~\cite{Heilmann-Lieb_71,Yang_89PRL,Yang-Zhang_90MPL}.
Essler, Korepin and Schoutens proved that,
when the system has the $so(4)$ symmetry,
the Bethe states are
$so(4)$-highest~\cite{Essler-Korepin_92aNPB}.
They also showed in a combinatorial way that all the 
eigenstates are obtained by taking the 
$so(4)$-descendants of the Bethe states
into account~\cite{Essler-Korepin_92bNPB}.
On the other hand the completeness of Bethe ansatz
for the system with odd sites, which has just
$sl(2)$ symmetry related to the spin degrees of freedom,
has not been discussed.  

In this article, we study the Bethe states
in the one-dimensional Hubbard model. In particular, 
we deal with the system on a ring 
with Aharonov-Bohm-type magnetic flux. 
The system has $so(4)$ symmetry
only at  special values of the AB-flux strength and
the $so(4)$ symmetry reduces to 
spin-$sl(2)$ symmetry for other values. More precisely, 
for a generic value of the AB-flux strength,  
the $so(4)$ symmetry breaks into the direct sum 
of the charge-$u(1)$ and the spin-$sl(2)$ symmetry.  
Varying the AB-flux strength,  
we investigate solutions of Lieb-Wu equations.  
Here we recall that all the enumeration of Bethe states 
we have mentioned above  
are based on the string hypothesis.
However, the violation of the string hypothesis is 
numerically observed: for the spin-$\frac{1}{2}$ isotropic 
Heisenberg spin chain, some of the string solutions 
reduce to real solutions when the number of sites is large, 
which is called redistribution phenomenon~\cite{%
Essler-Korepin_92JPA,Ilakovac_99PRB,Juttner-Dorfel_93JPA}.
Thus, without making any approximation, 
we discuss the existence of Bethe ansatz solutions with the AB-flux. 
In particular, we show the existence of solutions 
of the Lieb-Wu equations with  
an arbitrary number of up-spins and one down-spin. 
Here we employ a graphical 
approach~\cite{Deguchi_00PR,Nishino-Deguchi_03PRB}.
We exactly count the number of solutions in the case 
and verify that the enumeration with the string hypothesis
is correct only in the $so(4)$-case.
We find that the Lieb-Wu equations for the system with 
only the spin-$sl(2)$ symmetry have more solutions than 
those in the $so(4)$-case.
Next we study the combinatorial completeness of Bethe ansatz.
We obtain the relation among the characters of $so(4)$-modules 
through the power series identities similar to 
Kirillov's~\cite{Kirillov_85JSM,Kirillov_87JSM}, which gives a 
new proof of Essler-Korepin-Schoutens' combinatorial 
completeness of Bethe ansatz~\cite{Essler-Korepin_92bNPB}.
We also introduce a new combinatorial formula derived from
the relation among the characters of 
$\big(u(1)\oplus sl(2)\big)$-modules.
The formula suggests the combinatorial completeness of 
Bethe ansatz for the system only with the spin-$sl(2)$ 
symmetry, which has not been discussed in the literature.

The Bethe ansatz solutions with the AB-flux should be quite important 
in the low-dimensional physics of the one-dimensional Hubbard model. 
As pointed out by Kohn, 
the low frequency conductivity is directly 
related to the shift of the energy levels due to twisted boundary
conditions~\cite{Kohn_64PRA}.  
Sharpening Kohn's argument on electron systems in any dimensions, 
Shastry and Sutherland discussed effective mass 
of the one-dimensional Hubbard model through 
the twisted boundary conditions~\cite{Shastry-Sutherland_90PRL}.  
Furthermore, Kawakami and Yang obtained 
an explicit expression for the effective mass of the electric
conductivity for the one-dimensional Hubbard model 
\cite{Kawakami-Yang_90PRL,Kawakami-Yang_91PRB}. 
 For the Bethe ansatz solutions with the twisted boundary conditions, 
there are other aspects such as persistent current associated 
with the AB-flux. 

The article is organized as follows: 
in Section~\ref{sec:symmetry-Bethe},
we review the symmetry of the one-dimensional Hubbard 
model and the Bethe ansatz method.
We also describe how to construct eigenstates
other than the Bethe states.
In Section~\ref{sec:LW-eq_one-down} we prove the existence 
of solutions for the Lieb-Wu equations with one down-spin 
and exactly count the number of Bethe states 
in varying the strength of AB-flux.
In Section~\ref{sec:characters} we study the combinatorial
formulas for counting the Bethe states in terms of the characters 
of $so(4)$- and $\big(u(1)\oplus sl(2)\big)$-modules.
The final section is devoted to summary and concluding remarks.

%%%%%%%%%%%%%%%%%%%%%%%%%%%%%%%%%%%%%%%%%%%%%%%%%%%%%%%%%%%%%%%%%%%%%%
\section{Bethe ansatz method}
\label{sec:symmetry-Bethe}

\subsection{Hubbard model and Lieb-Wu equations}
We introduce the Hubbard model on an $L$-site ring
with Aharonov-Bohm-type magnetic flux $\Phi$.
Let $c_{is}^{\dagger}$ and $c_{is}, 
(i\in\mathbb{Z}/L\mathbb{Z}, 
s\in\{\uparrow,\downarrow\})$ be
the creation and annihilation operators of electrons satisfying
$\{c_{is},c_{jt}\}=\{c_{is}^{\dagger},c_{jt}^{\dagger}\}=0$ and
$\{c_{is},c_{jt}^{\dagger}\}=\delta_{ij}\delta_{st}$,
and define the number operators by 
$n_{is}\Define c_{is}^{\dagger}c_{is}$.
We consider the Fock space $V$ of electrons 
with the vacuum state $|0\rangle$ ($\dim V=4^{L}$).
The one-dimensional Hubbard model is described by the following 
Hamiltonian acting on $V$:
\begin{align}
\label{eq:Hubbard_Ham}
 &H_{\phi}
  =-\sum_{1\leq i\leq L}\sum_{s=\uparrow,\downarrow}
  (\e^{\i\phi}c_{is}^{\dagger}c_{i+1,s}
   +\e^{-\i\phi}c_{i+1,s}^{\dagger}c_{is})
  +U\sum_{1\leq i\leq L}\Big(n_{i\uparrow}-\frac{1}{2}\Big)
                  \Big(n_{i\downarrow}-\frac{1}{2}\Big),
\end{align}
where we assume 
$\phi\Define \Phi/L\in\mathbb{R}/2\pi\mathbb{R}$ and $U>0$.
It is clear that $H_{\phi}=H_{\phi+2\pi}$.
Furthermore $H_{\phi}$ has the same energy spectra as 
those of $H_{\phi+\frac{2\pi}{L}}$.
Hence we often restrict the region of $\phi$ 
to $0\leq\phi<\frac{2\pi}{L}$ in what follows.

Let $N$ be the number of electrons and $M$ that of down-spins.
We assume $0\leq 2M\leq N\leq L$.
Let $\{k_{i}|i\!=\!1,2,\ldots,N\}, 
(\mathrm{Re}(k_{i})\in\mathbb{R}/2\pi\mathbb{R})$
denote a set of wavenumbers of $N$ electrons and
$\{\lambda_{\alpha}|\alpha\!=\!1,2,\ldots,M\}$
that of rapidities of $M$ down-spins.
Given a set of spin configuration
$\{s_{i}|i\!=\!1,2,\ldots,N\}$
with $N-M$ up-spins and $M$ down-spins,
the Bethe state with $\{k_{i},\lambda_{\alpha}\}$ 
has the following form:
\begin{align}
\label{eq:Bethe-state}
 |k,\lambda;s\rangle_{N,M}^{\phi}
 =\sum_{\{1\leq x_{i}\leq L\}}
  \psi_{k,\lambda}(x;s)
  c_{x_{1},s_{1}}^{\dagger}c_{x_{2},s_{2}}^{\dagger}\cdots 
  c_{x_{N},s_{N}}^{\dagger}|0\rangle.
\end{align}
The coefficients $\psi_{k,\lambda}(x;s)$ 
in \eqref{eq:Bethe-state} are explicitly given 
in~\cite{Woynarovich_82JPC}.
The Bethe states~\eqref{eq:Bethe-state} are
eigenstates of the Hamiltonian~\eqref{eq:Hubbard_Ham}
if $\{k_{i},\lambda_{\alpha}\}$ satisfy the following equations:
\begin{align}
\label{eq:Lieb-Wu_eq}
 &\e^{\sqrt{-1}k_{i}L}=
  \prod_{1\leq\beta\leq M}
  \frac{\lambda_{\beta}-\sin(k_{i}\!+\!\phi)-\i U/4}
       {\lambda_{\beta}-\sin(k_{i}\!+\!\phi)+\i U/4}, \nn\\
 &\prod_{1\leq i\leq N}
  \frac{\lambda_{\alpha}-\sin(k_{i}\!+\!\phi)-\i U/4}
       {\lambda_{\alpha}-\sin(k_{i}\!+\!\phi)+\i U/4}
  =\prod_{\beta(\neq \alpha)}
   \frac{\lambda_{\alpha}-\lambda_{\beta}-\i U/2}
        {\lambda_{\alpha}-\lambda_{\beta}+\i U/2},
\end{align}
which are coupled nonlinear equations 
called Lieb-Wu equations~\cite{Lieb-Wu_68PRL}.
The Lieb-Wu equations have not been solved analytically.
However it predicts some important results on thermodynamic 
properties of the system through Takahashi's string 
hypothesis~\cite{Takahashi_72PTP,Takahashi_74PTP,Korepin-Essler%
,Takahashi_book}.
In terms of the solutions $\{k_{i},\lambda_{\alpha}\}$ 
of the Lieb-Wu equations~\eqref{eq:Lieb-Wu_eq}, 
energy eigenvalues are written as
\begin{align}
\label{eq:eigenvalues_cop}
 & H_{\phi} |k,\lambda;s\rangle_{N,M}^{\phi}
 =E|k,\lambda;s\rangle_{N,M}^{\phi},\quad
  E=-2\!\!\sum_{1\leq i\leq N}\!\!
  \cos(k_{i}\!+\!\phi)+\frac{1}{4}U(L\!-\!2N).
\end{align}
Recall that the Bethe states~\eqref{eq:Bethe-state} give
only the eigenstates with $0\leq 2M\leq N\leq L$.
In order to construct other eigenstates,
we need to consider the symmetries of the system. 

Hereafter we shall sometimes abbreviate the superscript 
$\phi$ in $|k,\lambda;s\rangle_{N,M}^{\phi}$. 

%%%%%%%%%%%%%%%%%%%%%%%%%%%%%%%%%%%%%%%%%%%%%%%%%%%%%%%%%%%%%%%%%%%%%%
\subsection{Symmetries}

The $U$-independent symmetries of the Hubbard model 
are classified in \cite{Heilmann-Lieb_71}.
First we review the symmetries connected with the spin 
and charge degrees of freedom~\cite{Yang_89PRL,Yang-Zhang_90MPL,%
Shiroishi-Wadati_97JPSJ,Shiroishi-Ujino_98JPA}. 
Define the following operators 
related to the spin degrees of freedom:
\begin{align}
\label{eq:spin-sl(2)}
 &S_{z}\Define
  \frac{1}{2}\sum_{1\leq i\leq L}
  (n_{i\uparrow}-n_{i\downarrow}),\quad
  S_{+}\Define
  \sum_{1\leq i\leq L}
  c_{i\uparrow}^{\dagger}c_{i\downarrow},\quad
  S_{-}\Define(S_{+})^{\dagger}.
\end{align}
They give a representation of the algebra $sl(2)$ 
on the Fock space $V$. 
Since all the operators $\{S_{z},S_{\pm}\}$ commute with 
$H_{\phi}$~\eqref{eq:Hubbard_Ham} 
for an arbitrary value of $\phi$, it is said that
the system has spin-$sl(2)$ symmetry.
One finds another representation of $sl(2)$ related 
to the charge degrees of freedom,
\begin{align}
\label{eq:charge-sl(2)}
 &\eta_{z}\Define
  \frac{1}{2}\sum_{1\leq i\leq L}
  (1-n_{i\uparrow}-n_{i\downarrow}),\quad
  \eta_{+}\Define
  \sum_{1\leq i\leq L}\e^{\i(2\phi+\pi)i}
  c_{i\downarrow}c_{i\uparrow},\quad
  \eta_{-}\Define(\eta_{+})^{\dagger}.
\end{align}
Note that all the operators in \eqref{eq:charge-sl(2)}
are commutative with $\{S_{z},S_{\pm}\}$ \eqref{eq:spin-sl(2)}.
It is easy to see that the operator $\eta_{z}$ also 
commutes with $H_{\phi}$ for an arbitrary value of $\phi$.
However other operators $\eta_{\pm}$ commute with 
$H_{\phi}$ only for the special values of $\phi$ satisfying
$\frac{L}{2}+\frac{L\phi}{\pi}\in\mathbb{Z}$.
Thus the system has charge-$sl(2)$ symmetry
if $\frac{L}{2}+\frac{L\phi}{\pi}\in\mathbb{Z}$,
and it reduces to charge-$u(1)$ symmetry given by $\eta_{z}$
for other values of $\phi$.
Combining the above two kinds of $sl(2)$ symmetries, 
we see that the system has $so(4)(\simeq sl(2)\oplus sl(2))$ 
symmetry if $\frac{L}{2}+\frac{L\phi}{\pi}\in\mathbb{Z}$,
and $u(1)\oplus sl(2)$ symmetry otherwise.
For simplicity, we call the former $so(4)$-case
and the latter $sl(2)$-case.
For the later discussion, 
we define the Casimir operators for each $sl(2)$,
\begin{align}
 \vecvar{\eta}^{2}\Define
  \frac{1}{2}\Big(\eta_{+}\eta_{-}+\eta_{-}\eta_{+}\Big)
  +\eta_{z}^{2},\quad
 \vecvar{S}^{2}\Define
  \frac{1}{2}\Big(S_{+}S_{-}+S_{-}S_{+}\Big)+S_{z}^{2},\nn
\end{align}
which are employed to see the dimension of 
each representation.

Next we introduce the following three operators:
\begin{align}
 &T_{\mathrm{s}}\Define
  \prod_{1\leq i\leq L}P_{i\uparrow;\,i\downarrow},\quad
  T_{\mathrm{ph}}\Define
  \prod_{1\leq i\leq L}\prod_{s=\uparrow,\downarrow}
  (c_{is}^{\dagger}+c_{is}),\quad
  T_{\mathrm{r}}\Define
  \prod_{1\leq i\leq \lfloor\frac{L-1}{2}\rfloor}
  \prod_{s=\uparrow,\downarrow}P_{is;\,L-i,s}, \nn
\end{align}
where $P_{is;\,jt}\Define
1\!-\!(c_{is}^{\dagger}\!-\!c_{jt}^{\dagger})
(c_{is}\!-\!c_{jt})$ and
$\lfloor x\rfloor$ denotes the greatest integer in $x$.
One notices that $T_{\mathrm{s}}$ induces a spin-reversal
transformation, $T_{\mathrm{ph}}$ a particle-hole 
transformation and $T_{\mathrm{r}}$ a reflection
of the lattice.
Direct calculation shows
\[
 T_{\mathrm{s}}^{-1}H_{\phi}T_{\mathrm{s}}=H_{\phi},\quad
 T_{\mathrm{ph}}^{-1}H_{\phi}T_{\mathrm{ph}}
 =H_{\pi-\phi},\quad
 T_{\mathrm{r}}^{-1}H_{\phi}T_{\mathrm{r}}=H_{-\phi}.
\]
For the system with even $L$, i.e., a bipartite lattice,
we also introduce
\[
  T_{\mathrm{b}}\Define
  \prod_{1\leq i\leq\frac{L}{2}}\prod_{s=\uparrow,\downarrow}
  (-1)^{n_{2i,s}},\quad
  T_{\mathrm{b}}^{-1}H_{\phi}T_{\mathrm{b}}=H_{\phi+\pi}.
\]
Combining these operations, we obtain the following transformation 
properties of the Hamiltonian 
$H_{\phi}$~\eqref{eq:Hubbard_Ham}:
\begin{align}
\label{eq:s-e-r-ph}
 &T_{\mathrm{s}}^{-1}H_{\phi}T_{\mathrm{s}}=H_{\phi},
  \quad\quad
  T_{\mathrm{ph}}^{-1}T_{\mathrm{r}}^{-1}
  H_{\phi}T_{\mathrm{r}}T_{\mathrm{ph}}
  =H_{\phi+\pi},\quad 
  \text{for even and odd $L$}, \nn\\
 &T_{\mathrm{ph}}^{-1}T_{\mathrm{r}}^{-1}T_{\mathrm{b}}^{-1}
  H_{\phi}T_{\mathrm{b}}T_{\mathrm{r}}T_{\mathrm{ph}}
  =H_{\phi},\quad \text{for even $L$}. 
\end{align}
One notices that for both even and odd $L$, 
the system has spin-reversal symmetry.
While the system with even $L$ has particle-hole 
symmetry brought by the transformation
$T_{\mathrm{b}}T_{\mathrm{r}}T_{\mathrm{ph}}$,
the system with odd $L$ does not have
particle-hole symmetry for generic values of $\phi$ 
except the special values satisfying 
$\frac{L}{2}+\frac{L\phi}{\pi}\in\mathbb{Z}$

%%%%%%%%%%%%%%%%%%%%%%%%%%%%%%%%%%%%%%%%%%%%%%%%%%%%%%%%%%%%%%%%%%%%%%
\subsection{Construction of non-Bethe states}
\label{sec:other-eigenstates}

\begin{figure}[t]
\begin{center}
\includegraphics[width=110mm,clip]{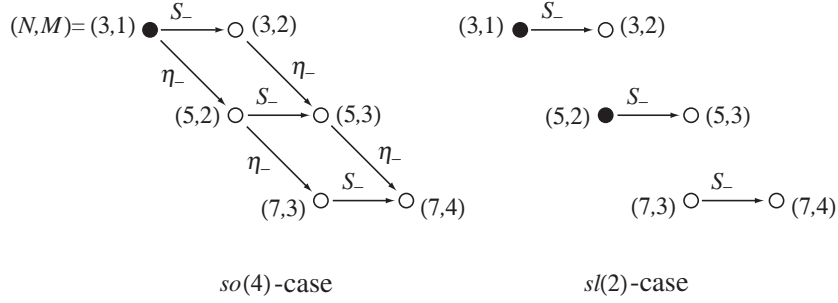}
\end{center}
\caption{Bethe states and non-Bethe states for $L=5$. 
The Bethe state with $(N,M)=(3,1)$
denoted by a closed circle produces five $so(4)$-descendants 
denoted by open circles if the system has $so(4)$ symmetry. 
As the $so(4)$ symmetry reduces to $sl(2)$ symmetry, 
the six eigenstates form three doublets of $sl(2)$.
Then we need one more Bethe state with $(5,2)$.
Note that the eigenstate with $(7,3)$ 
is not a Bethe states and is constructed through
the transformation $T_{\mathrm{r}}T_{\mathrm{ph}}$.}
\label{fig:so(4)-sl(2)}
\end{figure}

We construct eigenstates that are not included in 
the Bethe states \eqref{eq:Bethe-state}.
First we consider the system with $so(4)$ symmetry, i.e.,
both spin-$sl(2)$ and charge-$sl(2)$ symmetries.
Here we recall that the AB-flux parameter $\phi$ 
takes a special value: 
$\frac{L}{2}+\frac{L\phi}{\pi}\in\mathbb{Z}$ for even or odd $L$.  
Then, as shown in~\cite{Essler-Korepin_92aNPB} for even $L$ with 
$\phi=0$, the Bethe states~\eqref{eq:Bethe-state} 
characterized by finite-valued solutions of the Lieb-Wu 
equations~\eqref{eq:Lieb-Wu_eq} 
are $so(4)$-highest, 
i.e., $\eta_{+}|k,\lambda;s\rangle_{N,M}=
S_{+}|k,\lambda;s\rangle_{N,M}=0$.
Since
\begin{align}
\label{eq:S-eta}
 \vecvar{\eta}^{2}|k,\lambda;s\rangle_{N,M}
  =\eta(\eta+1)|k,\lambda;s\rangle_{N,M},\quad
 \vecvar{S}^{2}|k,\lambda;s\rangle_{N,M}
  =S(S+1)|k,\lambda;s\rangle_{N,M},
\end{align}
with $\eta=\frac{1}{2}(L\!-\!N)$ and $S=\frac{1}{2}(N\!-\!2M)$,
the Bethe state $|k,\lambda;s\rangle_{N,M}$ is the highest 
weight vector of an $(L\!-\!N\!+\!1)(N\!-\!2M\!+\!1)$-dimensional 
highest weight $so(4)$-module.
Here we note that $\eta +S$ is an integer for even $L$ 
\cite{Yang-Zhang_90MPL}, 
while it is a half-integer for odd $L$. 
The $so(4)$-descendant states of the Bethe state 
$|k,\lambda;s\rangle_{N,M}$,
\begin{align}
\label{eq:so(4)-descendants}
  (\eta_{-})^{n}(S_{-})^{m}|k,\lambda;s\rangle_{N,M},\quad
  (0<n\leq L-N, 0<m\leq N-2M),
\end{align}
are also eigenstates of 
$H_{\phi}$~\eqref{eq:Hubbard_Ham}
(see Fig.~\ref{fig:so(4)-sl(2)}).
They are energy degenerate
with $|k,\lambda;s\rangle_{N,M}$.
It is easy to see that such application of the 
lowering operators $\eta_{-}$ and $S_{-}$ to 
the Bethe states produces
other eigenstates than those with $0\leq 2M\leq N\leq L$.
Essler, Korepin and Schoutens counted the number of 
Bethe states~\eqref{eq:Bethe-state} and their 
$so(4)$-descendant states \eqref{eq:so(4)-descendants} 
under the string hypothesis to show the combinatorial 
completeness of Bethe ansatz~\cite{Essler-Korepin_92bNPB}.

Next we consider the system only with the charge-$u(1)$ and
the spin-$sl(2)$ symmetries.
The Bethe states~\eqref{eq:Bethe-state} are $sl(2)$-highest
and satisfy only the second 
relation in~\eqref{eq:S-eta}, which
means that the Bethe state $|k,\lambda;s\rangle_{N,M}^{\phi}$ 
is the highest weight vector of an 
$(N\!-\!2M\!+\!1)$-dimensional highest weight $sl(2)$-module.
The $sl(2)$-descendant states of the Bethe states
\begin{align}
\label{eq:sl(2)-descendant}
  (S_{-})^{m}|k,\lambda;s\rangle_{N,M}^{\phi},\quad
  (0<m\leq N-2M),
\end{align}
are eigenstates of $H_{\phi}$~\eqref{eq:Hubbard_Ham}
(see Fig.~\ref{fig:so(4)-sl(2)}).
One notices here that the Bethe states~\eqref{eq:Bethe-state} 
and their $sl(2)$-descendants~\eqref{eq:sl(2)-descendant}
do not produce the eigenstates with $L<N\leq 2L$ since
the application of the lowering operator $S_{-}$
does not change the number of electrons.
Such eigenstates with $L<N\leq 2L$ 
are constructed as follows: i)
by applying the transformation 
$T_{\mathrm{b}}T_{\mathrm{r}}T_{\mathrm{ph}}$ 
to the Bethe state $|k,\lambda;s\rangle_{2L-N,L-M}^{\phi},
(L<N\leq 2M\leq 2L)$
if $L$ is even,
or by applying $T_{\mathrm{r}}T_{\mathrm{ph}}$
to the Bethe state $|k,\lambda;s\rangle_{2L-N,L-M}^{\phi+\pi},
(L<N\leq 2M\leq 2L)$ 
obtained by the Lieb-Wu equations~\eqref{eq:Lieb-Wu_eq} 
with $\phi+\pi$ instead of $\phi$ if $L$ is odd,
we get the lowest weight vector of a highest weight 
$sl(2)$-module;
ii) the application of the raising operators
$(S_{+})^{n}, (0<n\leq 2M-N)$ 
to the lowest weight vector produces
other degenerate eigenstates.

Even if we are interested only in the system 
with $\phi=0$, the Bethe ansatz method needs the system
with $\phi=\pi$ for odd $L$.
One of the main purposes in this article is therefore to discuss 
whether or not the above procedure produces all the eigenstates.

%%%%%%%%%%%%%%%%%%%%%%%%%%%%%%%%%%%%%%%%%%%%%%%%%%%%%%%%%%%%%%%%%%%%%%
\section{Lieb-Wu equations with one down-spin}
\label{sec:LW-eq_one-down}

By employing a graphical approach,
we exactly count the number of finite-valued solutions for
the Lieb-Wu equations~\eqref{eq:Lieb-Wu_eq}
in the case when the system contains an arbitrary number
of up-spins and one down-spin, i.e.,
$N\geq 2$ and $M=1$.
We remark that such exact analysis is presented 
in~\cite{Deguchi_00PR,Nishino-Deguchi_03PRB} 
for the case $\phi=0$.
In this section,
we assume $0\leq \phi<\frac{2\pi}{L}$ since
$H_{\phi}$ has the same energy spectra as 
those of $H_{\phi+\frac{2\pi}{L}}$.
In the case $M=1$, the string hypothesis~\cite{Takahashi_72PTP}
predicts that two types of solutions exist;
one is the solution with only real wavenumbers $\{k_{i}\}$ and
another includes two complex wavenumbers.
We investigate such types of solutions below.

%%%%%%%%%%%%%%%%%%%%%%%%%%%%%%%%%%%%%%%%%%%%%%%%%%%%%%%%%%%%%%%%%%%%%%
\subsection{Real $k$ solutions}
\begin{figure}[t]
\begin{center}
\includegraphics[width=90mm,clip]{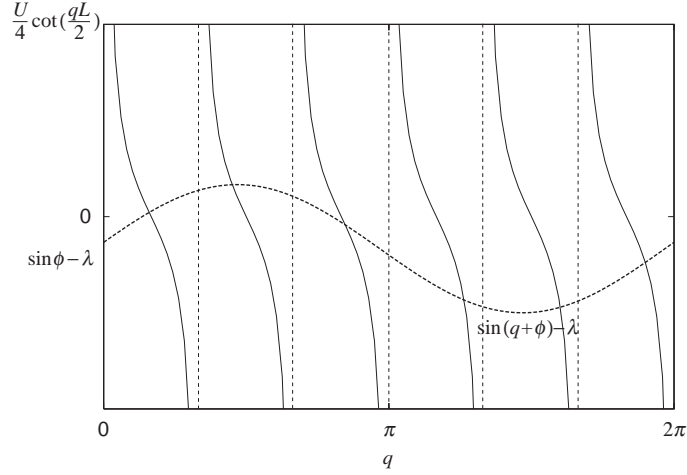}
\end{center}
\caption{The generic behaviour of 
$\sin(q\!+\!\phi)\!-\!\lambda$ 
and $\frac{U}{4}\cot(\frac{qL}{2})$ in \eqref{eq:LW_real-q}
under the condition $UL>8$ for $L=6$.
Horizontal dashed lines divide six branches
as specified by \eqref{eq:branch}.
Each branch has an intersection which leads to a solution
of \eqref{eq:LW_real-q}.}
\label{fig:sin-cot}
\end{figure}

First we consider the real solutions. 
For $M=1$, the Lieb-Wu equations~\eqref{eq:Lieb-Wu_eq} reduce to
\begin{align}
\label{eq:LW_1-down-spin}
 &\e^{\i k_{i}L}=
  \frac{\lambda-\sin(k_{i}\!+\!\phi)-\i U/4}
       {\lambda-\sin(k_{i}\!+\!\phi)+\i U/4},\quad(i=1,2,\ldots,N), \nn\\
 &\prod_{1\leq i\leq N}
  \frac{\lambda-\sin(k_{i}\!+\!\phi)-\i U/4}
       {\lambda-\sin(k_{i}\!+\!\phi)+\i U/4}=1. 
\end{align}
These are equivalent to the following equations:
\begin{align}
\label{eq:LW_real-k}
 \sin(k_{i}\!+\!\phi)-\lambda
 =\frac{U}{4}\cot\Big(\frac{k_{i}L}{2}\Big), \quad
 \exp\Big(\i\sum_{1\leq i\leq N}k_{i}L\Big)=1.
\end{align}
We investigate the real solutions 
for the first equation
\begin{align}
\label{eq:LW_real-q}
 \sin(q\!+\!\phi)-\lambda=\frac{U}{4}\cot\Big(\frac{qL}{2}\Big).
\end{align}
In the interval $0\leq q<2 \pi$,
its right hand side has $L$ branches 
\begin{align}
\label{eq:branch}
 \frac{2\pi}{L}\Big(\ell-\frac{1}{2}\Big)
 <q<
 \frac{2\pi}{L}\Big(\ell+\frac{1}{2}\Big),\quad
 \ell\in\Big\{\frac{2j-1}{2}\Big|j=1,2,\ldots,L\Big\}.
\end{align}
Note that, with a given $\ell$ satisfying \eqref{eq:branch},
the first equations in \eqref{eq:LW_1-down-spin}
are rewritten as
\[
  k_{i}L=2\pi\ell-
  2\arctan\Big(\frac{\lambda-\sin(k_{i}+\phi)}{U/4}\Big),
\]
which are convenient to relate
the $\ell$ to the (half-)integers appearing in
the string hypothesis~\cite{Takahashi_72PTP}.
By regarding $\lambda$ as a real parameter,
we seek a solution $q$ of \eqref{eq:LW_real-q} 
in one of the branches~\eqref{eq:branch}.
{}From the graphical discussion (Figure~\ref{fig:sin-cot}),
the solution in the branch $\ell$
is uniquely determined for arbitrary $\lambda$
under the following condition:
\[
 -1=\min_{0\leq q<2\pi}\,\frac{d}{dq}(\sin(q+\phi)-\lambda)
 >\max_{0\leq q<2\pi}\,\frac{d}{dq}
  \Big(\frac{U}{4}\cot\Big(\frac{qL}{2}\Big)\Big)
 =-\frac{UL}{8}.
\]

\noindent
Hence, for $U>\frac{8}{L}$, 
the solution of \eqref{eq:LW_real-q}
can be written as an increasing function of $\lambda$, i.e.,
$q=q_{\ell}(\lambda)$.
Given a non-repeating set $\{\ell_{i}|1\leq i\leq N\}\subset
\{\frac{2j-1}{2}|1\leq j\leq L\}$ of the branches,
the second equation in \eqref{eq:LW_real-k} is satisfied
when $\frac{L}{2\pi}\sum_{i}q_{\ell_{i}}(\lambda)\in\mathbb{Z}$.
The behaviour of the solution $q=q_{\ell}(\lambda)$
tells us that
\begin{align}
\label{eq:limit_q-lambda}
  \lim_{\lambda\to\pm\infty}\frac{L}{2\pi}
  \sum_{1\leq i\leq N}q_{\ell_{i}}(\lambda)
  =\sum_{1\leq i\leq N}\Big(\ell_{i}\pm\frac{1}{2}\Big).
\end{align}
Thus there exist $N-1$ values of $\lambda$ 
giving the following integer values
for $\frac{L}{2\pi}\sum_{i}q_{\ell_{i}}(\lambda)$:
\[
  m\in
  \Bigg\{\sum_{1\leq i\leq N}\!\Big(\ell_{i}-\frac{1}{2}\Big)+j
  \Bigg|j=1,2,\ldots,N-1\Bigg\}.
\]
Note that such $\{\lambda\}$ and integers $\{m\}$ are in
one-to-one correspondence due to 
$\frac{dq_{\ell}(\lambda)}{d\lambda}>0$.
As a consequence, 
the solutions $\{k_{i}=q_{\ell_{i}}(\lambda),\lambda\}$ 
are specified by the set of indices $\{\ell_{i},m\}$.
The number of possible $\{\ell_{i},m\}$ is given by
$\big({L \atop N}\big)(N-1)$.
Here we note that the number is consistent 
with the formula $Z(L;N,M)$ 
in~\cite{Takahashi_72PTP,Essler-Korepin_92bNPB}.

\begin{prop}
\label{prop:numb-real}
For $U>\frac{8}{L}$, the Lieb-Wu equations~\eqref{eq:Lieb-Wu_eq}
with $M=1$ have $\big({L \atop N}\big)(N-1)$ real 
solutions~\cite{Deguchi_00PR,Nishino-Deguchi_03PRB}.
\end{prop}

%%%%%%%%%%%%%%%%%%%%%%%%%%%%%%%%%%%%%%%%%%%%%%%%%%%%%%%%%%%%%%%%%%%%%%
\subsection{$k$-$\Lambda$-string solutions}
Next we consider the solutions including a couple
of complex wavenumbers.
We assume the form of solutions as
\[
  k_{i}\in\mathbb{R}/2\pi\mathbb{R},\;(i=1,2,\ldots,N-2),\quad
  k_{N-1}=\zeta\!-\!\i\xi,\quad
  k_{N}=\zeta\!+\!\i\xi,
\] 
where $0\leq \zeta<2\pi$ and $\xi>0$.
Note that $k_{N-1}$ and $k_{N}$ form a complex conjugate pair
which is referred to as $k$-$\Lambda$-2-string.
Then the first set of equations in \eqref{eq:LW_1-down-spin}
are rewritten in terms of real variables 
as follows  
\begin{subequations}
\label{eq:LW_string-k}
\begin{align}
\label{eq:LW_string-k_1}
 &\sin(k_{i}\!+\!\phi)-\lambda
  =\frac{U}{4}\cot\Big(\frac{k_{i}L}{2}\Big),\quad
 (i=1,2,\ldots,N-2), \\
\label{eq:LW_string-k_2}
 &\sin(\zeta\!+\!\phi)\cosh\xi-\lambda
  =\frac{U}{4}\frac{\sin(\zeta L)}{\cosh(\xi L)-\cos(\zeta L)}, \\
\label{eq:LW_string-k_3}
 &\cos(\zeta\!+\!\phi)\sinh\xi
  =-\frac{U}{4}\frac{\sinh(\xi L)}{\cosh(\xi L)-\cos(\zeta L)}.
\end{align}
\end{subequations}
On the other hand,
the second equation in~\eqref{eq:LW_1-down-spin} 
is equivalent to the following condition:
\begin{align}
\label{eq:LW_k-zeta}
 \sum_{1\leq i\leq N-2}k_{i}+2\zeta=\frac{2\pi}{L}m,\quad
 \text{with}\quad
 m=0,1,\ldots,NL-1.
\end{align}
In the same way as the previous case of section 3.2, 
if we consider a solution of each equation \eqref{eq:LW_string-k_1}
in one of the branches~\eqref{eq:branch},
and the solution can be written as a function of $\lambda$.
Given a set $\{\ell_{i}|1\leq i\leq N-2\}$ of non-repeating indices 
specifying the branches~\eqref{eq:branch}, 
we express the solutions of~\eqref{eq:LW_string-k_1}
as $k_{i}=q_{\ell_{i}}(\lambda), (1\leq i\leq N-2)$.
Then, from the relation \eqref{eq:LW_k-zeta}, 
the $\zeta$ is also written 
as a function of $\lambda$,
\begin{align}
\label{eq:zeta-lambda}
  \zeta=\zeta(\lambda)\Define\frac{\pi}{L}m
  -\frac{1}{2}\sum_{1\leq i\leq N-2}q_{\ell_{i}}(\lambda),
\end{align}
for fixed $\{\ell_{i}\}$ and $m$.

 For an illustration, we consider \eqref{eq:LW_string-k_2} and
\eqref{eq:LW_string-k_3} in the case $N=2$.
Since $\zeta$ does not depend on $\lambda$ in the case,
the equations \eqref{eq:LW_string-k_2} and
\eqref{eq:LW_string-k_3} decouple into the following: 
\begin{align}
 \label{eq:LW_string-k_N=2}
 &\lambda=\sin\Big(\frac{\pi}{L}m\!+\!\phi\Big)\cosh\xi, \quad
 \sinh\xi=-\frac{U}{4\cos(\frac{\pi}{L}m\!+\!\phi)}f^{(2)}(\xi),
\end{align}
where
\[
 f^{(2)}(\xi)\Define\frac{\sinh(\xi L)}{\cosh(\xi L)-(-1)^{m}}=
  \begin{cases}
   \coth(\xi L/2) &\text{for } m\in 2\mathbb{Z}, \\
   \tanh(\xi L/2) &\text{for } m\in 2\mathbb{Z}+1 . \\
  \end{cases}
\]
We seek a solution of the second equation 
 in \eqref{eq:LW_string-k_N=2}
through graphical discussion.
Since $f^{(2)}(\xi)>0$ when $\xi>0$, we need the condition 
$\frac{\pi}{2}<\frac{\pi}{L}m\!+\!\phi<\frac{3\pi}{2}$
so that the second equation of \eqref{eq:LW_string-k_N=2} 
has a solution.
If such $m$ is even, it is straightforward that
the second equation in \eqref{eq:LW_string-k_N=2}
determines a unique solution $\xi(>0)$
since $\lim_{\xi\to\infty}f^{(2)}(\xi)=1$.
For odd $m$, the equation has a unique solution
if the condition 
\[
 1=\frac{d(\sinh\xi)}{d\xi}(0)<
 \min_{m\in 2\mathbb{Z}+1 \atop 
 \frac{\pi}{2}<\frac{\pi}{L}m+\phi<\frac{3\pi}{2}}
 \Big(-\frac{U}{4\cos(\frac{\pi}{L}m\!+\!\phi)}
 \frac{df^{(2)}}{d\xi}(0)\Big)=\frac{UL}{8},
\]
is satisfied.
The number of allowed values of $m$  here depends on $L$ and $\phi$, 
\begin{align}
m\in
\begin{cases}
 \big\{\frac{L}{2}-\frac{L}{\pi}\phi+j\big|j=1,2,\ldots,L-1\big\}, 
 & \frac{L}{2}-\frac{L}{\pi}\phi\in\mathbb{Z}, \\[1mm]
 \big\{\big\lfloor\frac{L}{2}-\frac{L}{\pi}\phi\big\rfloor+j
       \big|j=1,2,\ldots,L\big\}, 
 & \frac{L}{2}-\frac{L}{\pi}\phi\notin\mathbb{Z}.
\end{cases}
\nn
\end{align}
Let $\xi_{m}^{(2)}$ denote the solution specified by the $m$.
By substituting the solution $\xi_{m}^{(2)}$ into
the first equation in \eqref{eq:LW_string-k_N=2},
one immediately obtains $\lambda$.
Recall that, in the cases of 
$\frac{L}{2}-\frac{L}{\pi}\phi\in\mathbb{Z}$, i.e.,
even $L$,(respectively, odd $L$) and $\phi=0$ or $\frac{\pi}{L}$, 
(respectively, $\phi=\frac{\pi}{2L}$ or $\frac{3\pi}{2L}$),
the system has the $so(4)$ symmetry.
Thus, as the $so(4)$ symmetry reduces to 
the spin-$sl(2)$ symmetry through the change of 
AB-flux strength $\phi$, 
the number of solutions for the Lieb-Wu equations increases.

\begin{figure}[t]
\begin{center}
\includegraphics[width=90mm,clip]{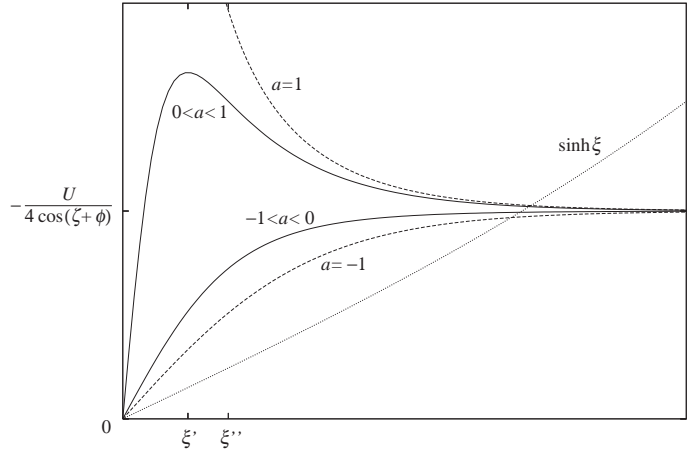}
\end{center}
\caption{The behaviour of $\sinh\xi$ and
$-\frac{U}{4\cos(\zeta+\phi)}f_{a}(\xi)$ in \eqref{eq:LW_xi-q_1}
under the conditions $UL>8$ and 
$\frac{\pi}{2}<\zeta+\phi<\frac{3\pi}{2}$.
The $\xi^{\prime}$ and $\xi^{\prime\prime}$ respectively denote 
the maximum and the turning point of 
$-\frac{U}{4\cos(\zeta+\phi)}f_{a}(\xi)$ in the cases $0<a<1$}
\label{fig:f_a(xi)}
\end{figure}

Let us consider the case $N>2$.
By inserting \eqref{eq:zeta-lambda} into \eqref{eq:LW_string-k_3},
we have
\begin{subequations}
\label{eq:LW_xi-q}
\begin{align}
\label{eq:LW_xi-q_1}
 &\sinh\xi=-\frac{U}{4\cos(\zeta(\lambda)\!+\!\phi)}
  f_{a}(\xi), \\
 &f_{a}(\xi)\Define
  \frac{\sinh(\xi L)}{\cosh(\xi L)-a},\quad\quad
  a=(-)^{m}\cos\Big(\frac{L}{2}
  \sum_{1\leq i\leq N-2}q_{\ell_{i}}(\lambda)\Big).
\end{align}
\end{subequations}
Note that $|a|\leq 1$.
One sees that, for a fixed $a$, the function $f_{a}(\xi)$ has 
the following properties: i) $f_{a}(0)=0$,
$\frac{df_{a}}{d\xi}(0)\geq\frac{L}{2}$, 
$\lim_{\xi\to\infty}f_{a}(\xi)=1$;
ii) if $a<0$, $f_{a}(\xi)$ is 
monotonically increasing and concave; 
iii) if $a>0$, $f_{a}(\xi)$ has
a single positive maximum at $\xi^{\prime}(>0)$ 
and a single turning point at $\xi^{\prime\prime}(>\xi^{\prime})$.
These properties are sufficient to 
discuss the solution $\xi$ of \eqref{eq:LW_xi-q} 
for arbitrary $\lambda$.
{}From the graphical discussion similar to the case $N=2$
(see Figure~\ref{fig:f_a(xi)}),
this determines a unique $\xi$ as a function of $\lambda$ 
under the condition
\[
 1=\frac{d(\sinh\xi)}{d\xi}(0)<
 \min_{|a|\leq 1\atop\frac{\pi}{2}<\zeta+\phi<\frac{3\pi}{2}}
 \Big(-\frac{U}{4\cos(\zeta\!+\!\phi)}
 \frac{df_{a}}{d\xi}(0)\Big)
 =\min_{|a|\leq 1}\frac{UL}{4(1-a)}=\frac{UL}{8},
\]
if and only if 
$\frac{\pi}{2}<\zeta(\lambda)+\phi=\frac{\pi}{L}m\!+\!\phi
 \!-\!\frac{1}{2}\sum_{i}q_{\ell_{i}}(\lambda)
 <\frac{3\pi}{2}$.
By using \eqref{eq:limit_q-lambda},
it is sufficient to have a unique solution for \eqref{eq:LW_xi-q}
that the integer $m$ satisfies
\[
  \sum_{1\leq i\leq N-2}
  \!\!\Big(\ell_{i}+\frac{1}{2}\Big)+\frac{L}{2}
  -\frac{L}{\pi}\phi
  <m<
  \sum_{1\leq i\leq N-2}
  \!\!\Big(\ell_{i}-\frac{1}{2}\Big)+\frac{3L}{2}
  -\frac{L}{\pi}\phi,
\]
that is, 
\begin{align}
\label{eq:string_m}
  m\in
\begin{cases}
 \Big\{\sum_{i}\Big(\ell_{i}+\frac{1}{2}\Big)
       +\frac{L}{2}-\frac{L}{\pi}\phi+j
       \Big|j=1,2,\ldots,L\!-\!N\!+\!1\Big\}, 
 & \frac{L}{2}-\frac{L}{\pi}\phi\in\mathbb{Z}, \\[2mm]
 \Big\{\sum_{i}\Big(\ell_{i}+\frac{1}{2}\Big)
       +\big\lfloor\frac{L}{2}-\frac{L}{\pi}\phi\big\rfloor+j
       \Big|j=1,2,\ldots,L\!-\!N\!+\!2\Big\}, 
 & \frac{L}{2}-\frac{L}{\pi}\phi\notin\mathbb{Z}.
\end{cases}
\end{align}
Note that $\lim_{\lambda\to\pm\infty}\xi(\lambda)
=\xi_{m-\sum_{i}(\ell_{i}\pm\frac{1}{2})}^{(2)}$, which
is well-defined for the above $m$.
We see that,
for the values of $m$ given in \eqref{eq:string_m},
the equation \eqref{eq:LW_string-k_2} with 
$\xi(\lambda)$ and $\zeta(\lambda)$ 
\begin{align}
\label{eq:LW_string-k_3_1}
 \lambda
 &=\sin\big({\textstyle
 \frac{\pi}{L}m\!+\!\phi\!
 -\!\frac{1}{2}\sum_{i}q_{\ell_{i}}(\lambda)}\big)
  \cosh\big(\xi(\lambda)\big)
 -\frac{U}{4}\frac{\sin(\frac{L}{2}\sum_{i}q_{\ell_{i}}(\lambda))}
 {\cos\big(\frac{L}{2}\sum_{i}q_{\ell_{i}}(\lambda)\big)
  -(-)^{m}\cosh\big(\xi(\lambda) L\big)} \nn\\
 &\definE g\big(\{q_{\ell_{i}}(\lambda)\},\xi(\lambda)\big),
\end{align}
determines $\lambda$.
In fact, since $q_{\ell_{i}}(\lambda)$ and $\xi(\lambda)$
are continuous functions of $\lambda$ and the function $g$
satisfies the following: 
\[
 \lim_{\lambda\to\pm\infty}
 g\big(\{q_{\ell_{i}}(\lambda)\},\xi(\lambda)\big)
 =g\Big({\textstyle\{
  \frac{2\pi}{L}(\ell_{i}\pm\frac{1}{2})\},
  \xi_{m-\sum_{i}(\ell_{i}\pm\frac{1}{2})}^{(2)}}\Big) . 
\]
Here $g$ is a continuous and finite function 
with respect to $\lambda$.
Hence there exists a solution $\lambda$ 
for the equation \eqref{eq:LW_string-k_3_1}.

\begin{prop}
\label{prop:numb-string}
For $U>\frac{8}{L}$,
the Lieb-Wu equations~\eqref{eq:Lieb-Wu_eq} with $M=1$ have
$({L \atop N-2})(L-N+1)$ $k$-$\Lambda$-2-string solutions 
if the system has the $so(4)$ symmetry, 
and they have
$({L \atop N-2})(L-N+2)$ $k$-$\Lambda$-2-solutions, otherwise.
\end{prop}

One notices that, only for the system with 
the $so(4)$ symmetry, 
the number of $k$-$\Lambda$-2-solutions is consistent 
with the string hypothesis~\cite{Takahashi_72PTP}.
Note that, for $0<U<\frac{8}{L}$, some of
the $k$-$\Lambda$-2-strings may disappear. 
In Appendix~\ref{sec:redistribution}, 
we numerically investigate the case $N=2$
and show that, for $0<U<\frac{8}{L}$, 
the $k$-$\Lambda$-2-strings with odd $m$
disappear, while additional
real solutions appear~\cite{Essler-Korepin_92aNPB}.
For the system with only the spin-$sl(2)$
symmetry, the Lieb-Wu equations have
more $k$-$\Lambda$-2-solutions than those
expected by the string hypothesis.

%%%%%%%%%%%%%%%%%%%%%%%%%%%%%%%%%%%%%%%%%%%%%%%%%%%%%%%%%%%%%%%%%%%%%%
\subsection{Completeness of Bethe ansatz for $L=3$}
\label{sec:L=3}

Applying the above results,
we now show that all the eigenstates
can be constructed through the Bethe ansatz method in a simple case.
We consider the case $L=3$ and 
$\phi\neq \frac{\pi}{2L},\frac{3\pi}{2L}$ 
when the system does not have
$so(4)$ symmetry but 
the spin-$sl(2)$ symmetry.
We note that the completeness of eigenstates in this situation
has not been discussed in the literature.

The number of Bethe states $|k,\lambda;s\rangle_{N,M}^{\phi},
(0\leq 2M\leq N\leq 3)$ is exactly calculated as follows:
the case $M=0$ is trivial since the eigenstates 
are those of lattice free fermion system;
for the cases $(N,M)=(2,1)$ and $(3,1)$, 
we have obtained the following formulas from
Proposition~\ref{prop:numb-real} 
and \ref{prop:numb-string}:
\[
\text{$\sharp$(Bethe states)}=
\begin{cases}
 \displaystyle{
 \left({3 \atop N}\right)\left({N\!-\!1 \atop 1}\right)
 }
 & \text{ for real solutions},\\[4mm]
 \displaystyle{
 \left({3 \atop N\!-\!2}\right)
 \left({5\!-\!N \atop 1}\right)
 }
 & \text{ for $k$-$\Lambda$-2-string solutions}.
\end{cases}
\]
Since each Bethe state $|k,\lambda;s\rangle_{N,M}^{\phi}$
corresponds to the highest weight vector
of a highest weight $sl(2)$-module,
we should count their $sl(2)$-descendant states 
\[
 (S_{-})^{n}|k,\lambda;s\rangle_{N,M}^{\phi}, 
 \quad\quad (0<n\leq N-2M).
\]
The eigenstates with $4\leq N\leq 6$, 
which are not Bethe states nor 
their $sl(2)$-descendant states, 
are constructed through the
transformation $T_{\mathrm{r}}T_{\mathrm{ph}}$
as we have described in Section~\ref{sec:other-eigenstates}.
Indeed, by applying $T_{\mathrm{r}}T_{\mathrm{ph}}$
to the Bethe state $|k,\lambda;s\rangle_{6-N,3-M}^{\phi+\pi},
(4\leq N\leq 2M\leq 6)$ of the system described 
by the Hamiltonian $H_{\phi+\pi}$, we get the eigenstate 
$T_{\mathrm{r}}T_{\mathrm{ph}}
|k,\lambda;s\rangle_{6-N,3-M}^{\phi+\pi}$
of $H_{\phi}$ with $4\leq N\leq 2M\leq 6$ 
which is the lowest weight vector
of a highest weight $sl(2)$-module.
We also count the eigenstates 
\[
 (S_{+})^{n}T_{\mathrm{r}}T_{\mathrm{ph}}
 |k,\lambda;s\rangle_{6-N,3-M}^{\phi+\pi},
 \quad\quad (0<n\leq 2M-N).
\]
Table~\ref{tab:count-L=3} indeed shows that we obtain
$64=4^{3}=\mathrm{dim}V$ eigenstates, 
which give a complete system of the Fock space $V$.

\begin{table}[h]
\def\arraystretch{1.2}
\begin{center}
\begin{tabular}{|cc|cc|c|c|c||c|}
\hline
 $N$ & $M$ & $6\!-\!N$ & $3\!-\!M$
 & type of solutions
 & $\sharp(\text{Bethe})$ 
 & $sl(2)$ sym. 
 & $\sharp(\text{state})$ \\
\hline
\hline
 0 & 0 &&& real & 1 & 1 & 1 \\
 1 & 0 &&& real & 3 & 2 & 6 \\
 2 & 0 &&& real & 3 & 3 & 9 \\
 2 & 1 &&& real & 3 & 1 & 3 \\
 2 & 1 &&& $k$-$\Lambda$-2-string & 3 & 1 & 3 \\
 3 & 0 &&& real & 1 & 4 & 4 \\
 3 & 1 &&& real & 2 & 2 & 4 \\
 3 & 1 &&& $k$-$\Lambda$-2-string & 6 & 2 & 12 \\
\hline
 4 & 2 & 2 & 1 & real & 3 & 1 & 3 \\
 4 & 2 & 2 & 1 & $k$-$\Lambda$-2-string & 3 & 1 & 3 \\
 4 & 3 & 2 & 0 & real & 3 & 3 & 9 \\
 5 & 3 & 1 & 0 & real & 3 & 2 & 6 \\
 6 & 3 & 0 & 0 & real & 1 & 1 & 1 \\
\hline
\hline
 &&&&&&& 64 \\
\hline
\end{tabular}
\end{center}
\label{tab:count-L=3}
\caption{Enumeration of eigenstates for $L=3$ 
and $\phi\neq \frac{\pi}{2L},\frac{3\pi}{2L}$.}
\end{table}

%%%%%%%%%%%%%%%%%%%%%%%%%%%%%%%%%%%%%%%%%%%%%%%%%%%%%%%%%%%%%%%%%%%%%%
\section{Combinatorial completeness of Bethe ansatz}
\label{sec:characters}

The Bethe ansatz method was first introduced in the case of 
one-dimensional spin-$\frac{1}{2}$ isotropic Heisenberg 
spin chain~\cite{Bethe_31ZP}.
Bethe assumed the string hypothesis and estimated the number 
$Z(N;M)$ of solutions for the Bethe equations with $M$ down-spins
on an $N$-site chain as
\begin{align}
\label{eq:comb-XXX}
 &Z(N;M)=\!\!\!
  \sum_{\{M_{n}\} \atop M=\sum nM_{n}}\!\!
  \prod_{m\geqslant 1}
  \left({P_{m}\!+\!M_{m}\atop M_{m}}\right),
\end{align}
where $M_{n}$ denotes the number of $n$-strings
composing a solution and $P_{n}=N-2M+2\sum_{m(>n)}(m-n)M_{m}$.
He obtained the following summation formula:
\begin{align}
\label{eq:sum-XXX}
  Z(N;M)=\Big({N \atop M}\Big)
  -\Big({N \atop M-1}\Big).
\end{align}
which implies that the number $Z(N;M)$ of Bethe states 
is interpreted as the multiplicity of 
$(N\!-\!2M\!+\!1)$-dimensional irreducible $sl(2)$-modules 
in the tensor product of $N$
two-dimensional irreducible $sl(2)$-modules.
By taking into account that the Bethe states are 
$sl(2)$-highest and generate $(N-2M+1)$
$sl(2)$-descendant states,
the completeness of Bethe ansatz is shown
in a combinatorial way as
\[
  \sum_{0\leq M\leq\lfloor N/2\rfloor}
  (N-2M+1)Z(N;M)=2^{N}.
\]
where $\lfloor x\rfloor$ denotes the greatest integer in $x$.

It is known that, in general, 
solutions of Bethe equations do 
not have the nature assumed in the
string hypothesis~\cite{Essler-Korepin_92JPA,%
Ilakovac_99PRB,Juttner-Dorfel_93JPA}.
Indeed, for $N>21$, some of the $2$-string solutions are 
redistributed to real solutions that are not counted 
in $Z(N;M)$~\cite{Essler-Korepin_92JPA}.
Hence, when we actually employ the the formula $Z(N;M)$ 
to show the completeness of Bethe ansatz, 
we must regard such redistributed 
real solutions as $2$-string solutions.

We apply the techniques developed 
in~\cite{Kirillov_85JSM,Kirillov_87JSM}
to the Hubbard model with $so(4)$ symmetry.
Indeed a new proof for Essler-Korepin-Schoutens' combinatorial 
completeness of Bethe ansatz~\cite{Essler-Korepin_92bNPB}
is obtained as a corollary of the relation among 
the characters of $so(4)$-modules.
Moreover, based on the results in the previous section,
we propose 
the conjectural formula related to the combinatorial 
completeness of Bethe ansatz for the system
with only the charge-$u(1)$ and spin-$sl(2)$ symmetry.
In both cases, we obtain the formulas corresponding
to \eqref{eq:sum-XXX}, which has not been established
even for the $so(4)$-case.

%%%%%%%%%%%%%%%%%%%%%%%%%%%%%%%%%%%%%%%%%%%%%%%%%%%%%%%%%%%%%%%%%%%%%%
\subsection{Kirillov's power series and $Q$-system}
First we give three lemmas introduced in the case of
one-dimensional isotropic Heisenberg spin 
chain~\cite{Kirillov_85JSM,Kirillov_87JSM}.
Detailed proofs are given 
in~\cite{Kirillov_85JSM,Kirillov_87JSM}.
Let $a_{1},a_{2},\ldots,a_{l}$ be a set of integers
for $l>\!\!>L$.
Define a set of formal power series 
$\{\varphi_{n}(z_{n},z_{n+1},\ldots,z_{l})|1\leq n\leq l\}$ by
\begin{align}
 &\psi_{n}(z)\Define (1-z)^{-a_{n}+2M-1},\nn\\
 &\varphi_{n}(z_{n},z_{n+1},\ldots,z_{l})\Define
  \psi_{n}(z_{n})\varphi_{n+1}((1\!-\!z_{n})^{-2}z_{n+1},\ldots,
                               (1\!-\!z_{n})^{-2(l-n)}z_{l}). \nn
\end{align}
\begin{lm}
\label{lm:expand-phi}
The power series $\varphi_{n}(z_{n},\ldots,z_{l})$ 
has the following expression:
\[
  \varphi_{n}(z_{n},\ldots,z_{l})
  =\sum_{M_{n},\ldots,M_{l}\geq 0}
  \prod_{n\leq m\leq l}
  \left({\mathcal{P}_{m}(a_{m})\!+\!M_{m} \atop M_{m}}\right)
  z_{n}^{M_{n}}\cdots z_{l}^{M_{l}},\quad\quad
  \text{for}\quad 1\leq n\leq l,
\]
where
\[
  \mathcal{P}_{n}(a_{n})
  =a_{n}\!-\!2M\!+\!2\sum_{m>n}(m\!-\!n)M_{m}.
\]
\end{lm}
\begin{proof}
The case $n=l$ is given by the formula,
\[
  (1-z)^{-\alpha-1}
  =\sum_{m\geq 0}
   \left({\alpha\!+\!m \atop m}\right)z^{m}.
\]
Then the case $1\leq n<l$ is proved by induction on $n$.
\end{proof}

Introduce the variables 
$\{z_{n}^{(k)}|1\leq n\leq l,0\leq k\leq n\}$ through
\[
  z_{n}^{(0)}=z_{n},\quad\quad
  z_{n}^{(k)}=(1-z_{k}^{(k-1)})^{-2(n-k)}z_{n}^{(k-1)},\quad
  \text{for}\quad 1\leq k\leq n.
\]
\begin{lm}
\label{lm:phi-z}
The power series $\varphi_{1}(z_{1},\ldots,z_{l})$ is rewritten 
in terms of the variables $\{z_{n}^{(k)}\}$,
\[
  \varphi_{1}(z_{1},\ldots,z_{l})
  =\prod_{1\leq n\leq l}(1-z_{n}^{(n-1)})^{-a_{n}+2M-1}.
\]
\end{lm}
\begin{proof}
Since
\begin{align}
 \varphi_{k}(z_{k}^{(k-1)},\ldots,z_{l}^{(k-1)}) 
 &=\psi_{k}(z_{k}^{(k-1)})
   \varphi_{k+1}((1\!-\!z_{k}^{(k-1)})^{-2}z_{k+1}^{(k-1)},
          \ldots,(1\!-\!z_{k}^{(k-1)})^{-2(l-k)}z_{l}^{(k-1)}) \nn\\
 &=(1-z_{k}^{(k-1)})^{-a_{k}+2M-1}
   \varphi_{k+1}(z_{k+1}^{(k)},\ldots,z_{l}^{(k)}), \nn
\end{align}
for $1\leq k\leq l-1$, the lemma is proved.
\end{proof}

Define the polynomials $\{Q_{n}=Q_{n}(t)|n\in\mathbb{Z}_{\geq 0}\}$ 
through the recursion relation,
\begin{align}
\label{eq:recursion-Q}
  Q_{n+2}=Q_{n+1}-tQ_{n},\quad Q_{0}=Q_{1}=1.
\end{align}

\begin{lm}
\label{lm:relations-Q}
\begin{subequations}
\begin{align}
 \label{eq:Q-system}
  \text{i)}&\quad 
  Q_{n}^{2}=Q_{n+1}Q_{n-1}+t^{n},\quad \text{for}\quad n\geq 1,
  \\
 \label{eq:phi-Q}
  \text{ii)}&\quad
  \varphi_{1}(t,t^{2},\ldots,t^{l})
  =\prod_{1\leq n\leq l}(Q_{n+1}Q_{n}^{-2}Q_{n-1})^{-a_{n}+2M-1}, \\
  \text{iii)}&\quad
  Q_{n}(t(v))=\frac{(1-v)^{n-1}(1-v^{n+1})}{(1-v^{2})^{n}},\quad
  \text{where}\quad t(v)\Define\frac{v}{(1+v)^{2}}.
\end{align}
\end{subequations}
\end{lm}
\begin{proof}
i) Use induction on $n$.

\noindent
ii) Set $z_{n}^{(0)}=z_{n}=t^{n}, (1\leq n\leq l)$ 
in $\{z_{n}^{(k)}\}$. One obtains the following relations:
\[
  1-z_{k}^{(k-1)}=Q_{k+1}Q_{k}^{2}Q_{k-1},\quad
  z_{n}^{(k)}=Q_{k+1}^{-2(n-k)}Q_{k}^{2(n-k-1)}t^{n},\quad
  \text{for}\quad 1\leq k\leq n.
\]
Combining these with Lemma~\ref{lm:phi-z}, 
we can prove \eqref{eq:phi-Q}.

\noindent
iii) One can directly verify that the $Q_{n}(t(v))$ satisfies 
the recursion relation~\eqref{eq:recursion-Q} with $t=t(v)$.
\end{proof}
The relations~\eqref{eq:Q-system} are called 
the $Q$-system of type $sl(2)$, 
which is a key object in~\cite{Kirillov_85JSM}.
Indeed the expression \eqref{eq:phi-Q} produces
an identity among the characters of $sl(2)$-modules.
The $Q$-system also plays a significant role
in the combinatorial identities associated with the 
XXZ-Heisenberg spin chain and 
its generalizations~\cite{Kirillov-Liskova_97JPA,%
Kuniba-Nakanishi_00PC,Kuniba-Nakanishi_02JA,%
Kuniba-Nakanishi-Tsuboi_02LMP}.

%%%%%%%%%%%%%%%%%%%%%%%%%%%%%%%%%%%%%%%%%%%%%%%%%%%%%%%%%%%%%%%%%%%%%%
\subsection{Combinatorial formulas}

Using the above lemmas, we discuss the Hubbard-case.
In the similar way, we define 
$\varphi_{n}^{\prime}(z_{n},z_{n+1},\ldots,z_{l})$ 
and $\mathcal{P}_{n}^{\prime}(a_{n}^{\prime})$ 
by replacing $a_{n}$ and $2M$
with $a_{n}^{\prime}$ and $N$ 
in $\varphi_{n}(z_{n},z_{n+1},\ldots,z_{l})$ and 
$\mathcal{P}_{n}(a_{n})$, respectively.
Define 
\[
  \varphi(s,t)
  \Define (1+s)^{L}
  \varphi_{1}^{\prime}(s^{2}t,s^{4}t^{2},\ldots,s^{2l}t^{l})
  \varphi_{1}(t,t^{2},\ldots,t^{l}).
\]
The following is straightforward from Lemma~\ref{lm:expand-phi}:
\begin{align}
 \varphi(s,t)
  =\hspace{-3mm}
  \sum_{0\leq N_{r}\leq L\atop \{M_{n},M_{n}^{\prime}\}}
  \hspace{-3mm} 
  \left({L \atop N_{r}}\right)\!\!\!
  \prod_{1\leq n\leq l}\!\!
  \left({\mathcal{P}_{n}^{\prime}(a_{n}^{\prime})\!
         +\!M_{n}^{\prime}
         \atop M_{n}^{\prime}}\right)\!\!
  \left({\mathcal{P}_{n}(a_{n})\!+\!M_{n} \atop M_{n}}\right)\!
  s^{N_{r}+2\sum_{m\geq 1}mM_{m}^{\prime}}\,
  t^{\sum_{m\geq 1}m(M_{m}+M_{m}^{\prime})}. \nn
\end{align}
Then the coefficient of $s^{N}t^{M}, (0\leq 2M\leq N\leq L)$ 
in $\varphi(s,t)$ is expressed by
\[
  \mathcal{Z}(L,\{a_{n},a_{n}^{\prime}\};N,M)
  \Define\hspace{-3mm}
  \sum_{{\{N_{r},M_{n},M_{n}^{\prime}\} \atop
        {N=N_{r}+2\sum n M_{n}^{\prime} \atop 
         M=\sum n(M_{n}+M_{n}^{\prime})}}}
  \hspace{-3mm} 
  \left({L \atop N_{r}}\right)\!\!
  \prod_{1\leq n\leq l}\!\!
  \left({\mathcal{P}_{n}^{\prime}(a_{n}^{\prime})\!
         +\!M_{n}^{\prime}
         \atop M_{n}^{\prime}}\right)\!
  \left({\mathcal{P}_{n}(a_{n})\!+\!M_{n} \atop M_{n}}\right).
\]
where the sum runs over all configurations 
$\{N_{r},M_{n},M_{n}^{\prime}\geq 0\}$
such that $N=N_{r}+2\sum_{n\geq 0} n M_{n}^{\prime}$ and
$M=\sum_{n\geqslant 1}n(M_{n}+M_{n}^{\prime})$.
We calculate explicit forms for the power series $\varphi(s,t)$ 
after taking the special values of $\{a_{n}\}$ 
and $\{a_{n}^{\prime}\}$. 

Introduce
\begin{subequations}
\begin{align}
\label{eq:Z-so(4)}
 &Z(L;N,M)=\hspace{-7mm}
  \sum_{{\{N_{r},M_{n},M_{n}^{\prime}\} \atop
        {N=N_{r}+2\sum n M_{n}^{\prime} \atop 
         M=\sum n(M_{n}+M_{n}^{\prime})}}}
  \hspace{-5mm}
  \left({L \atop N_{r}}\right)\prod_{n\geqslant 1}
  \left({P_{n}^{\prime}\!+\!M_{n}^{\prime} 
         \atop M_{n}^{\prime}}\right)
  \left({P_{n}\!+\!M_{n} \atop M_{n}}\right), \\
\label{eq:Z-sl(2)}
 &\widetilde{Z}(L;N,M)=\hspace{-7mm}
  \sum_{{\{N_{r},M_{n},M_{n}^{\prime}\} \atop
        {N=N_{r}+2\sum n M_{n}^{\prime} \atop 
         M=\sum n(M_{n}+M_{n}^{\prime})}}}
  \hspace{-5mm}
  \left({L \atop N_{r}}\right)\prod_{n\geqslant 1}
  \left({P_{n}^{\prime}\!+\!M_{n}^{\prime}\!+\!n
         \atop M_{n}^{\prime}}\right)
  \left({P_{n}\!+\!M_{n} \atop M_{n}}\right), 
\end{align}
\end{subequations}
where
\begin{align}
 P_{n}^{\prime}=
 L\!-\!N\!+\!2\sum_{m>n}(m\!-\!n)M_{m}^{\prime},\quad\quad
 P_{n}=N\!-\!2M\!+\!2\sum_{m>n}(m\!-\!n)M_{m}. \nn
\end{align} 
Note that $P_{n}^{\prime},P_{n}\geq 0$ 
due to $0\leq 2M\leq N\leq L$.

The $Z(L;N,M)$~\eqref{eq:Z-so(4)}
is the very number of Bethe states
for the Hubbard model estimated under the string 
hypothesis~\cite{Takahashi_72PTP,Takahashi_book,%
Essler-Korepin_92bNPB}. 
In terms of the string hypothesis, 
$N_{r}$ denotes the number of real $k$'s, 
$M_{n}$ the number of $\Lambda$-$n$-strings, 
and $M_{n}^{\prime}$ the number of $k$-$\Lambda$-$2n$-strings.  
We have verified in the previous section 
that, if the system has $so(4)$ symmetry, the number
\[
 Z(L;N,1)
 =\left({L \atop N}\right)\left({N\!-\!1 \atop 1}\right)
 +\left({L \atop N\!-\!2}\right)
  \left({L\!-\!N\!+\!1 \atop 1}\right),
\]
gives the correct number of solutions
for the Lieb-Wu equations~\eqref{eq:Lieb-Wu_eq} with $M=1$.
In fact the first and second terms in the above $Z(L;N,1)$ 
correspond to the following two cases: $N$ real $k$'s 
($N_{r}=N$) with a $\Lambda$-$1$-string ($M_{1}=1$),  
and $N-2$ real $k$'s ($N_{r}=N-2$) with 
a $k$-$\Lambda$-2-string ($M_{1}^{\prime}=1$), 
respectively.

We now propose the 
$\widetilde{Z}(L;N,M)$ \eqref{eq:Z-sl(2)} 
as a formula counting the number of Bethe states 
for the system with charge-$u(1)$ and spin-$sl(2)$ symmetries.
Indeed 
\[
 \Tilde{Z}(L;N,1)
 =\left({L \atop N}\right)\left({N\!-\!1 \atop 1}\right)
 +\left({L \atop N\!-\!2}\right)
  \left({L\!-\!N\!+\!2 \atop 1}\right),
\]
is consistent with the number and the string-type of solutions
for the Lieb-Wu equations~\eqref{eq:Lieb-Wu_eq} with $M=1$
in the $sl(2)$-case. The first term corresponds to the case
of $N$ real $k$'s  ($N_{r}=N$) with 
a $\Lambda$-1-string ($M_{1}=1$),  
and the second term to the case of $N-2$ real $k$'s ($N_{r}=N-2$) 
with a $k$-$\Lambda$-2-string ($M_{1}^{\prime}=1$). 
We note that, for $L<N\leq 2L$, we interpret 
$\widetilde{Z}(L;N,M)$ as the number of the lowest weight 
vectors $T_{\mathrm{r}}T_{\mathrm{ph}}
|k,\lambda;s\rangle_{2L-N,L-M}^{\phi+\pi}$.
To derive $\Tilde{Z}(L;N,M)$ \eqref{eq:Z-sl(2)}
for $M\geq 2$ from Takahashi's string center 
equations~\cite{Takahashi_72PTP,Takahashi_book},
we need to appropriately extend the region 
of the allowed (half-)integers characterizing the Bethe states.

In both the $so(4)$- and $sl(2)$-cases, 
one must also take a redistribution phenomenon 
into consideration~\cite{Essler-Korepin_92bNPB,%
Essler-Korepin_92JPA,Ilakovac_99PRB,Juttner-Dorfel_93JPA};
what it means here will be more clear in 
Appendix~\ref{sec:redistribution}.

\begin{prop}
\label{prop:main-identities}
We have the following identities:
\begin{align}
\label{eq:identities}
 \text{i)}\quad
 &(1\!+\!u)^{L}(1\!+\!uv)^{L}(1\!-\!u^{2}v)(1\!-\!v)
 =\sum_{0\leq N\leq 2L+2 \atop 0\leq M\leq L+1}
  Z(L;N,M)u^{N}v^{M}, \nn\\
 \text{ii)}\quad
 &(1\!+\!u)^{L}(1\!+\!uv)^{L}(1\!-\!v)
 =\sum_{0\leq N\leq 2L \atop 0\leq M\leq L+1}
  \widetilde{Z}(L;N,M)u^{N}v^{M}.
\end{align}
\end{prop}
\begin{proof}
By using Lemma~\ref{lm:relations-Q},
we have
\[
  \varphi(s,t)
  =(1+s)^{L}\prod_{1\leq n\leq l}
   (Q_{n+1}^{\prime}Q_{n}^{\prime\, -2}
    Q_{n-1}^{\prime})^{-a_{n}^{\prime}+N-1}
   (Q_{n+1}Q_{n}^{-2}Q_{n-1})^{-a_{n}+2M-1},
\]
where $Q_{n}^{\prime}(t)\Define Q_{n}(s^{2}t)$.
Through the change of variables
\[
  s=\frac{u(1+v)}{1+u^{2}v},\quad
  t=\frac{v}{(1+v)^{2}},\quad
  ds\,dt=\frac{(1-u^{2}v)(1-v)}{(1+u^{2}v)^{2}(1+v)^{2}}du\,dv,
\]
the coefficient of $s^{N}t^{M}$ 
in $\varphi(s,t)$ is calculated as
\begin{align}
&\mathcal{Z}(L,\{a_{n},a_{n}^{\prime}\};N,M)
 =\Res_{s=0,t=0}\;\varphi(s,t)
  \frac{ds}{s^{N+1}}\frac{dt}{t^{M+1}} \nn\\
&=\Res_{u=0,v=0}\Big(
  \frac{(1+u)^{L}(1+uv)^{L}}{(1+u^{2}v)^{L}}\!\!
  \prod_{1\leq n\leq l}\!\!
  (Q_{n+1}^{\prime}Q_{n}^{\prime\, -2}
   Q_{n-1}^{\prime})^{-a_{n}^{\prime}+N-1}
  (Q_{n+1}Q_{n}^{-2}Q_{n-1})^{-a_{n}+2M-1} \nn\\
&\quad\quad\times
  \frac{(1+u^{2}v)^{N+1}}{(1+v)^{N+1}}(1+v)^{2M+2}
  \frac{(1-u^{2}v)(1-v)}{(1+u^{2}v)^{2}(1+v)^{2}}\Big)
  \frac{du}{u^{N+1}}\frac{dv}{v^{M+1}} \nn\\
&=\Res_{u=0,v=0}\Big(
  \frac{(1+u)^{L}(1+uv)^{L}}{(1+u^{2}v)^{L}}
  \frac{(1-u^{2}v)(1-v)}{(1+v)^{N}}\!\!
  \prod_{1\leq n\leq l}\!\!\!
  (1-u^{2n}v^{n})^{b_{n}^{\prime}}(1-v^{n})^{b_{n}}\Big)
  \frac{du}{u^{N+1}}\frac{dv}{v^{M+1}}, \nn
\end{align}
where we have introduced
\[
  b_{n}=-a_{n}+2a_{n-1}-a_{n-2},\quad\quad
  b_{n}^{\prime}=-a_{n}^{\prime}
  +2a_{n-1}^{\prime}-a_{n-2}^{\prime},\quad\quad
  (1\leq n\leq l),
\]
with $a_{-1}\!=\!a_{0}\!=a_{-1}^{\prime}\!=\!a_{0}^{\prime}\!=\!0$.
Note that, in the third equality, we have employed
the assumption $l>\!\!>L$.
Setting $a_{n\geq 1}=L$ and $a_{n\geq 1}^{\prime}=N$, we have 
$-b_{1}^{\prime}=b_{2}^{\prime}=L, -b_{1}=b_{2}=N$ and
$b_{n\geq 3}^{\prime}=b_{n\geq 3}^{\prime}=0$,
which proves the identity i) in~\eqref{eq:identities}.
And, setting $a_{n\geq 1}=L+n$ and $a_{n\geq 1}^{\prime}=N$, 
we have $b_{1}^{\prime}=-L-1,b_{2}^{\prime}=L, -b_{1}=b_{2}=N$
and $b_{n\geq 3}^{\prime}=b_{n\geq 3}^{\prime}=0$,
which proves the identity ii).
\end{proof}

Through the change of variables $u=xy^{-1}$ and $v=y^{2}$,
we have
\begin{subequations}
\begin{align}
\label{eq:xy-id_so(4)}
 &(x\!+\!x^{-1}\!+\!y\!+\!y^{-1})^{L}(x\!-\!x^{-1})(y\!-\!y^{-1})
 =\sum_{0\leq N\leq 2L+2 \atop 0\leq M\leq L+1}
  Z(L;N,M)x^{-L+N-1}y^{-N+2M-1}, \\
\label{eq:xy-id_sl(2)}
 &(x\!+\!x^{-1}\!+\!y\!+\!y^{-1})^{L}(y\!-\!y^{-1})
 =\sum_{0\leq N\leq 2L \atop 0\leq M\leq L+1}
  \widetilde{Z}(L;N,M)x^{-L+N}y^{-N+2M-1}. 
\end{align}
\label{eq:xy-identities}
\end{subequations}
First we consider the relation~\eqref{eq:xy-id_so(4)}.
If we express the left-hand side of \eqref{eq:xy-id_so(4)} 
as $F(x,y)$, we find the property 
$F(x^{-1},y)=F(x,y^{-1})=-F(x,y)$.
Then the first relation~\eqref{eq:xy-id_so(4)} 
is rewritten as
\begin{align}
\label{eq:so(4)-identity}
  (x\!+\!x^{-1}\!+\!y\!+\!y^{-1})^{L}=\!\!\!
  \sum_{0\leq N\leq L\atop 
  0\leq M\leq\lfloor N/2\rfloor}\!\!\!Z(L;N,M)
  \frac{x^{L-N+1}\!-\!x^{-L+N-1}}{x-x^{-1}}
  \frac{y^{N-2M+1}\!-\!y^{-N+2M-1}}{y-y^{-1}},
\end{align}
where $\lfloor x\rfloor$ denotes the greatest integer in $x$.
Let $V_{n}^{\prime}$ be an $(n\!+\!1)$-dimensional irreducible 
$sl(2)$-module associated with the 
charge-$sl(2)$ symmetry
and $V_{n}$ that associated with 
the spin-$sl(2)$ symmetry.
We introduce an $(n\!+\!1)(m\!+\!1)$-dimensional irreducible 
$so(4)$-module by the tensor product
$V_{n,m}=V_{n}^{\prime}\otimes V_{m}$. 
Through the representation \eqref{eq:spin-sl(2)} and 
\eqref{eq:charge-sl(2)} of $so(4)$, 
the Fock space $V$ of the $L$-site system
is isomorphic to the tensor product
$(V_{1,0}\oplus V_{0,1})^{\otimes L}$ as an $so(4)$-module.
Note that
\[
  \vecvar{\eta}^{2}|v\rangle
  =\frac{n}{2}\left(\frac{n}{2}+1\right)|v\rangle,
  \quad
  \vecvar{S}^{2}|v\rangle
  =\frac{m}{2}\left(\frac{m}{2}+1\right)|v\rangle,
  \quad\quad
  \text{for } |v\rangle\in V_{n,m}\subset V.
\]
We now decompose the Fock space $V$ into 
the direct sum of $V_{n,m}$.
{}From the relations~\eqref{eq:S-eta}, each Bethe state 
$|k,\lambda;s\rangle_{N,M}$ corresponds to 
the highest weight vector belonging to
$V_{L-N,N-2M}=V_{2\eta,2S}$.
The characters of the $so(4)$-module $V_{n,m}$ 
are calculated as
\[
  \ch V_{n,m}
  =\frac{x^{n+1}-x^{-n-1}}{x-x^{-1}}
   \frac{y^{m+1}-y^{-m-1}}{y-y^{-1}}.
\]
To be precise, $x=\e^{\Lambda_{1}^{\prime}}$
and $y=\e^{\Lambda_{1}}$ where both $\Lambda_{1}^{\prime}$
and $\Lambda_{1}$ are the fundamental weight of $sl(2)$ 
and they are orthogonal to each other.
One notices that, in terms of the characters, 
the identity~\eqref{eq:so(4)-identity} can be rewritten as
\begin{align}
\label{eq:ch-so4}
  (\ch V_{1,0}+\ch V_{0,1})^{L}=\!\!\!
  \sum_{0\leq N\leq L\atop 
  0\leq M\leq\lfloor N/2\rfloor}\!\!\!Z(L;N,M)
  \ch V_{L-N,N-2M}. 
\end{align}

\begin{thm}[Multiplicity formula]
\label{thm:multiplicity_so4}
The multiplicity of the irreducible component $V_{L-N,N-2M}$ 
in the tensor product $(V_{1,0}\oplus V_{0,1})^{\otimes L}$
is given by $Z(L;N,M)$,
\begin{align}
\label{eq:decomposition_so4}
  (V_{1,0}\oplus V_{0,1})^{\otimes L}=\hspace{-4mm}
  \bigoplus_{0\leq N\leq L\atop 
  0\leq M\leq\lfloor N/2\rfloor}\hspace{-4mm}
  Z(L;N,M) V_{L-N,N-2M}.
\end{align}
\end{thm}

Next we turn to the relation \eqref{eq:xy-id_sl(2)}.
If we express the left-hand side of \eqref{eq:xy-id_sl(2)} as 
$\widetilde{F}(x,y)$, we find
$\widetilde{F}(x,y^{-1})=-\widetilde{F}(x,y)$.
This gives
\begin{align}
\label{eq:sl(2)-identity}
  \big(1\!+\!x(y\!+\!y^{-1})\!+\!x^{2}\big)^{L}=\!\!\!
  \sum_{0\leq N\leq 2L\atop 
  0\leq M\leq\lfloor N/2\rfloor}\!\!\!
  \widetilde{Z}(L;N,M)\,x^{N}\,
  \frac{y^{N-2M+1}\!-\!y^{-N+2M-1}}{y-y^{-1}}.
\end{align}
Let $V^{(N)}\subset V$ be a subspace of the Fock space $V$
with $N$ electrons and let $V^{(N)}_{m}\subset V^{(N)}$ be an 
$(m\!+\!1)$-dimensional irreducible $sl(2)$-module
related to the spin-$sl(2)$ symmetry in $V^{(N)}$.
The $V^{(N)}_{m}$ can be regarded as a 
$1\!\times\!(m\!+\!1)$-dimensional 
irreducible $\big(u(1)\oplus sl(2)\big)$-module
by employing the representation \eqref{eq:spin-sl(2)} 
of $sl(2)$ and taking $u(1)=\mathbb{C}(L-2\eta_{z})$ 
with \eqref{eq:charge-sl(2)}.
Here
\[
 (L-2\eta_{z})|v\rangle
 =N|v\rangle,\quad
 \vecvar{S}^{2}|v\rangle
  =\frac{m}{2}\left(\frac{m}{2}+1\right)|v\rangle,
  \quad\quad
  \text{for } |v\rangle\in V_{m}^{(N)}\subset V.
\]
The Fock space $V$ of the $L$-site Hubbard model 
is isomorphic to the tensor product
$(V_{0}^{(0)}\oplus V_{1}^{(1)}\oplus V_{0}^{(2)})^{\otimes L}$ 
as a $\big(u(1)\oplus sl(2)\big)$-module.
We decompose the Fock space $V$ into the direct sum
of $V^{(N)}_{m}$.
{}From the first relation in~\eqref{eq:S-eta}, 
each Bethe state $|k,\lambda;s\rangle_{N,M}$ is
the highest weight vector belonging to
$V^{(N)}_{N-2M}=V^{(N)}_{2S}$.
The characters of $V_{m}^{(N)}$ are calculated as
\[
  \ch V_{m}^{(N)}
  =x^{N}\,\frac{y^{m+1}-y^{-m-1}}{y-y^{-1}}.
\]
In terms of the characters, 
the identity~\eqref{eq:sl(2)-identity} can be rewritten as
\begin{align}
\label{eq:ch-sl2}
  (\ch V_{0}^{(0)}+\ch V_{1}^{(1)}+\ch V_{0}^{(2)})^{L}=\!\!\!
  \sum_{0\leq N\leq 2L\atop 
  0\leq M\leq\lfloor N/2\rfloor}\!\!\!Z(L;N,M)
  \ch V_{N-2M}^{(N)}. 
\end{align}
\begin{thm}[Multiplicity formula]
\label{thm:multiplicity_sl2}
The multiplicity of the irreducible component $V_{N-2M}^{(N)}$ 
in the tensor product 
$(V_{0}^{(0)}\oplus V_{1}^{(1)}\oplus V_{0}^{(2)})^{\otimes L}$
is given by $\widetilde{Z}(L;N,M)$,
\begin{align}
\label{eq:decomposition_sl2}
  (V_{0}^{(0)}\oplus V_{1}^{(1)}\oplus V_{0}^{(2)})^{\otimes L}
  =\hspace{-4mm}
  \bigoplus_{0\leq N\leq 2L\atop 
  0\leq M\leq\lfloor N/2\rfloor}\hspace{-4mm}
  \widetilde{Z}(L;N,M) V_{N-2M}^{(N)}.
\end{align}
\end{thm}

\begin{cor}[Combinatorial completeness]
\label{cor:completeness}
We have
\begin{align}
 \text{i)}\quad
 &\dim V=\!\!\!
  \sum_{0\leq N\leq L\atop 
  0\leq M\leq\lfloor N/2\rfloor}\!\!\!
  (L\!-\!N\!+\!1)(N\!-\!2M\!+\!1)Z(L;N,M),\nn\\
 \text{ii)}\quad
 &\dim V=\!\!\!
  \sum_{0\leq N\leq 2L\atop 
  0\leq M\leq\lfloor N/2\rfloor}\!\!\!
  (N\!-\!2M\!+\!1)\widetilde{Z}(L;N,M).
\end{align}
\end{cor}
\begin{proof}
Consider the limit $x,y\to 1$ in the 
identities~\eqref{eq:so(4)-identity} and \eqref{eq:sl(2)-identity}.
\end{proof}
The identity i) in Corollary~\ref{cor:completeness}
reproduces the combinatorial completeness of Bethe states
for the Hubbard model with $so(4)$ symmetry 
obtained by Essler, Korepin and 
Schoutens~\cite{Essler-Korepin_92bNPB}.
The factor $(L\!-\!N\!+\!1)(N\!-\!2M\!+\!1)$ in i) 
corresponds to the dimension of the highest weight 
$so(4)$-module $V_{L-N,N-2M}$ with the highest weight 
vector $|k,\lambda;s\rangle_{N,M}$.

In Essler-Korepin-Schoutens' proof of 
i) in Corollary~\ref{cor:completeness}, they take
the sum on $N$ after taking that on $M$.
In our proof, the sums on $N$ and $M$ are taken
``simultaneously'' in the level of characters.

The factor $(N\!-\!2M\!+\!1)$ of the identity ii) in 
Corollary~\ref{cor:completeness} for $0\leq N\leq L$ is
the dimension of the highest weight
$\big(u(1)\oplus sl(2)\big)$-module $V_{N-2M}^{(N)}$
with the highest weight vector $|k,\lambda;s\rangle_{N,M}$.
For $L< N\leq 2L$, the factor $(N\!-\!2M\!+\!1)$  
should be interpreted as the dimension of the highest weight
$\big(u(1)\oplus sl(2)\big)$-module $V_{N-2M}^{(N)}$
with the lowest weight vector $T_{\mathrm{r}}T_{\mathrm{ph}}
|k,\lambda;s\rangle_{2L-N,L-M}^{\phi+\pi}$.
If even $L$, the identity ii) can be rewritten as
\[
  \dim V=\!\!\!
  \sum_{0\leq N\leq L\atop 
  0\leq M\leq\lfloor N/2\rfloor}\!\!\!
  2^{1-\delta_{N,L}}
  (N\!-\!2M\!+\!1)\widetilde{Z}(L;N,M),
\]
by considering the particle-hole symmetry of the system
\eqref{eq:s-e-r-ph}.
Thus we speculate that the identity ii) 
in Corollary~\ref{cor:completeness} accounts for
the combinatorial completeness of Bethe states
for the system with the charge-$u(1)$ and
the spin-$sl(2)$ symmetries.

The identities \eqref{eq:xy-identities} 
also enabled us to get the explicit form
of $Z(L;N,M)$ through the binomial theorem.
\begin{cor}
We obtain the summation formulas for $Z(L;N,M)$ 
and $\widetilde{Z}(L;N,M)$,
\begin{align}
\label{eq:summation}
 \text{i)}\quad
 &Z(L;N,M)=
  \left(L\!+\!2\atop M\right)\!
  \left(L\atop N\!-\!M\right)
  -\left(L\atop M\!-\!1\right)\!
   \left(L\!+\!2\atop N\!-\!M\!+\!1\right), \nn\\
 \text{ii)}\quad
 &\widetilde{Z}(L;N,M)=
 \left(L\atop M\right)\!
  \left(L\atop N\!-\!M\right)
  -\left(L\atop M\!-\!1\right)\!
  \left(L\atop N\!-\!M\!+\!1\right).
\end{align}
\end{cor}

%%%%%%%%%%%%%%%%%%%%%%%%%%%%%%%%%%%%%%%%%%%%%%%%%%%%%%%%%%%%%%%%%%%%%%
\section{Summary and concluding remarks}

In the framework of Bethe ansatz,
we have studied the Hubbard model with the AB-flux
that controls the symmetry of the system.
In Section~\ref{sec:LW-eq_one-down}
we have shown the existence of solutions 
for Lieb-Wu equations with an arbitrary 
number of up-spins and one down-spin.
We have found that the number of $k$-$\Lambda$-2-solutions 
increases as the $so(4)$ symmetry reduces to the spin-$sl(2)$ 
symmetry (Proposition \ref{prop:numb-string}).
The number of Bethe states is consistent with
the string hypothesis only in the $so(4)$-case.
In Section~\ref{sec:characters}
we have investigated the combinatorial formulas
giving the combinatorial completeness of Bethe states.
We have shown that the number of Bethe states can be 
interpreted as the multiplicity of irreducible components
in the tensor products of $so(4)$-modules
(Theorem~\ref{thm:multiplicity_so4}).
Essler-Korepin-Schoutens' combinatorial formula 
is reproduced by the relation~\eqref{eq:ch-so4}
among the characters of $so(4)$-modules 
(Corollary~\ref{cor:completeness}).
An advantage of our approach is that we can obtain
the summation formula~\eqref{eq:summation} for $Z(L;N,M)$.
We have also proposed a new combinatorial formula derived 
from the relation~\eqref{eq:ch-sl2} among the characters of 
$\big(u(1)\oplus sl(2)\big)$-modules.
The formula is related to the combinatorial 
completeness of Bethe ansatz in the $sl(2)$-case
(Corollary~\ref{cor:completeness}).
It should be remarked that, 
in Section~\ref{sec:LW-eq_one-down}, we have not 
proved the uniqueness of solutions.
Although the Lieb-Wu equations may have
other solutions that we have not expected,
the combinatorial formulas introduced in 
Section~\ref{sec:characters} give an evidence that, 
for $M=1$, we have found solutions enough 
to verify the combinatorial completeness of Bethe ansatz.
The problem is open for $M\geq 2$.

The combinatorial completeness of Bethe ansatz
has not been discussed for the one-dimensional 
isotropic Heisenberg spin chain with 
twisted boundary conditions.
Kirillov's identity~\cite{Kirillov_85JSM} also produces
the following formula:
\begin{align}
 &\Tilde{Z}(N;M)=
  \sum_{\{M_{n}\} \atop M=\sum nM_{n}}
  \prod_{n\geqslant 1}
  \left({P_{n}\!+\!M_{n}\!+\!n \atop M_{n}}\right)
  =\Big({N \atop M}\Big), \nn
\end{align}
where $P_{n}=N-2M+2\sum_{m(>n)}(m-n)M_{m}$.
It is expected that, if the redistribution phenomenon
of solutions for Bethe equations
\cite{Essler-Korepin_92JPA,Ilakovac_99PRB,Juttner-Dorfel_93JPA} 
is taken into consideration,
the formula corresponds
to the system with twisted boundary conditions.
We remark that the formula also appears 
in the different context~\cite{Kuniba-Okado_03RIMS}.

\section*{Acknowledgements}
The authors would like to thank Prof.~A.~Kuniba and Dr.~T.~Takagi 
for valuable discussion.
They also would like to thank 
Prof.~M.~Wadati and Dr.~M.~Shiroishi for helpful comments.
One of the authors (AN) appreciates the Research Fellowships of the
Japan Society for the Promotion of Science for Young Scientists. 
The present study is partially supported by the Grant-in-Aid for 
Encouragement of Young Scientists (A): No. 14702012. 

%%%%%%%%%%%%%%%%%%%%%%%%%%%%%%%%%%%%%%%%%%%%%%%%%%%%%%%%%%%%%%%%%%%%%%
\appendix

%%%%%%%%%%%%%%%%%%%%%%%%%%%%%%%%%%%%%%%%%%%%%%%%%%%%%%%%%%%%%%%%%%%%%%
\section{Redistribution phenomenon}
\label{sec:redistribution}

In Section~\ref{sec:LW-eq_one-down},
we exactly show the existence of solutions
of Lieb-Wu equations with $M=1$ for $U>\frac{8}{L}$.
There, real solutions have been specified by non-repeating indices
$\{\ell_{i}|i=1,2,\ldots,N\}$ and $m$.
But, for $0<U<\frac{8}{L}$, 
real solutions with repeating indices may appear 
at the same time as $k$-$\Lambda$-2-string solutions
disappear. 
Such redistribution of type of solutions is observed
in the isotropic Heisenberg model for a large number of sites 
\cite{Essler-Korepin_92JPA,Ilakovac_99PRB,Juttner-Dorfel_93JPA}.
Here we numerically investigate 
such phenomenon for the Hubbard model with $L=20$
and $N=2$~\cite{Essler-Korepin_92bNPB}.

\begin{figure}[h]
\begin{center}
\includegraphics[width=90mm,clip]{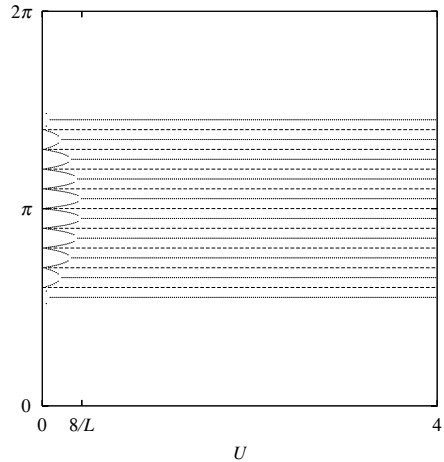}
\vspace{-5mm}
\end{center}
\caption{The real parts of solutions for 
Lieb-Wu equations with $L\!=\!20$ and $N\!=\!2$. 
The dashed lines correspond to 
the centers $\zeta$ of $k$-$\Lambda$-2-string solutions 
with even $m$, and the dots express the centers $\zeta$
of $k$-$\Lambda$-2-string solutions with odd $m$
and their redistribute real solutions.}
\label{fig:redist_re}
\end{figure}

\begin{figure}[h]
\begin{center}
\hspace{5mm}
\includegraphics[width=80mm,clip]{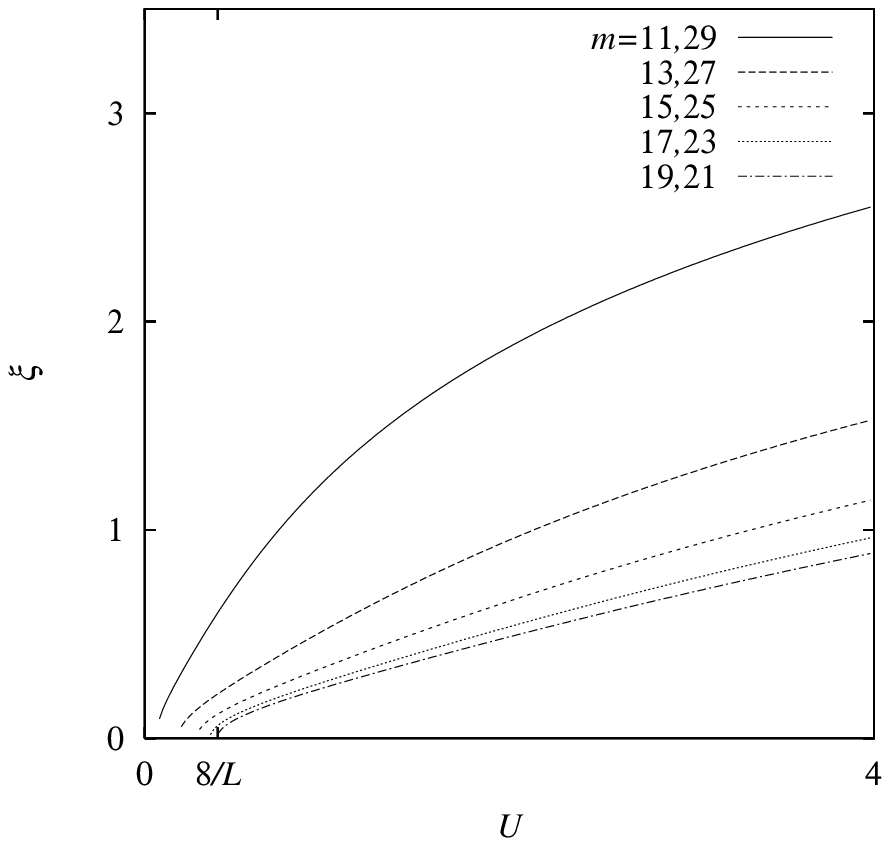}
\hspace{-15mm}
\includegraphics[width=80mm,clip]{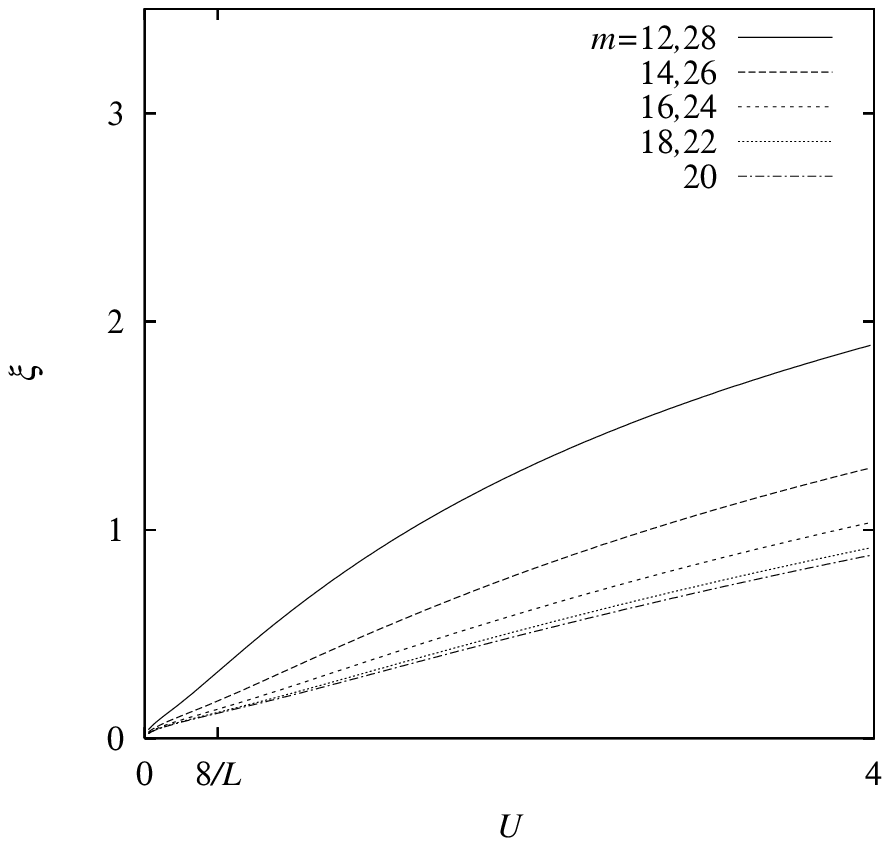}
\vspace{-5mm}
\end{center}
\caption{The imaginary parts $\xi$ of solutions 
for Lieb-Wu equations with $L\!=\!20$ and $N\!=\!2$.}
\label{fig:redist_img}
\end{figure}

As we have already mentioned,
$k$-$\Lambda$-2-string solutions with odd $m$ may disappear
for $0<U<\frac{8}{L}$ (see Figure~\ref{fig:f_a(xi)}).
For each $m$, the critical value of $U$ is exactly given 
by~\cite{Essler-Korepin_92bNPB},
\[
  U^{(m)}=-\frac{8}{L}\cos\Big(\frac{\pi}{L}m\Big)
  <\frac{8}{L}.
\]
Plotted on Figure~\ref{fig:redist_re} are the center 
$\zeta=\frac{\pi}{L}m, (m=11,12,\ldots,29)$ 
of $k$-$\Lambda$-2-string solutions 
and redistributed real solutions $\{k_{1},k_{2}\}$ 
for $L=20$ and $N=2$ in varying the value of $U$.
Plotted on Figure~\ref{fig:redist_img} are
their imaginary parts.
The $k$-$\Lambda$-2-string solutions with odd $m$ disappear 
for $U<U^{(m)}$ and, at the same time, real solutions with 
repeating indices $\{\ell_{1},\ell_{2};m\}
=\{\frac{m}{2},\frac{m}{2};m\}$ appear.
Note that, as $U\to 0$, all the $k$-$\Lambda$-2-string 
solutions on Figure~\ref{fig:redist_re} approach to 
the wavenumbers of lattice free fermion system.
Thus, when we apply the combinatorial formulas
in Corollary \ref{cor:completeness} to the case,
we must count the number of Bethe states by
regarding the real solutions with repeating indices
as $k$-$\Lambda$-string solutions.

%\bibliographystyle{amsplain}
%\bibliography{physref2}
\ifx\undefined\bysame
\newcommand{\bysame}{\leavevmode\hbox to3em{\hrulefill}\,}
\fi

\end{document}